\documentclass[traditabstract]{aa}  

\usepackage{graphicx}
\usepackage{natbib}
\usepackage{rotating}
\usepackage{txfonts}
\usepackage{color}
\usepackage{url}

\newcommand{\Kepler} {\textit{Kepler}\xspace}

\newcommand{\Hermes} {\textsc{Hermes}\xspace}
\newcommand{\hermes} {\textsc{Hermes}\xspace}
\newcommand{\sindex}{$\mathcal{S}$-index\xspace}
\newcommand{\Sindex}{$\mathcal{S}$-index\xspace}
\newcommand{\Ssymbol}{$\mathcal{S}$\xspace}

\usepackage[switch]{lineno} % line number
\begin{document}

   \title{Photospheric and chromospheric magnetic activity\\
   of seismic solar analogs}
   \subtitle{Observational inputs on the solar/stellar connection from \emph{Kepler} and \Hermes\thanks{Based on observations collected by the NASA {\it Kepler} space telescope and the \Hermes spectrograph mounted on the 1.2\,m \textsc{Mercator} telescope at the Spanish Observatorio del Roque de los Muchachos of the Instituto de Astrof\'isica de Canarias.}}
  
   \author{D. Salabert  \inst{1}
        \and
         R.~A. Garc\'ia \inst{1}
        \and
         P.~G. Beck \inst{1}
     	\and
     	R. Egeland \inst{2,3}
	\and
	P.~L. Pall\'e\inst{4,5}
	  \and   
         S. Mathur \inst{6}
         \and 
        T.~S. Metcalfe \inst{6,7}
           \and 
           \\
    	J.-D. do Nascimento Jr.\inst{8,9}
           \and
         T. Ceillier \inst{1}
     	\and
	M.~F. Andersen\inst{10}
	\and
	A. Trivi\~no Hage\inst{4,5} 
 }  
   \institute{Laboratoire AIM, CEA/DRF-CNRS, Universit\'e Paris 7 Diderot, IRFU/SAp, Centre de Saclay, 91191 Gif-sur-Yvette, France\\
              \email{david.salabert@cea.fr}
            \and
	High Altitude Observatory, National Center for Atmospheric Research, PO Box 3000, Boulder, CO 80307-3000, USA
	\and
	Department of Physics, Montana State University, Bozeman, MT 59717-3840, USA
	 \and
   Instituto de Astrof\'isica de Canarias,  E-38200 La Laguna, Tenerife, Spain
	     \and
      	Departamento de Astrof\'isica, Universidad de La Laguna, E-38205 La Laguna, Tenerife, Spain  
         \and
 	 Space Science Institute, 4750 Walnut street Suite\#205, Boulder, CO 80301, USA	
	 \and
	 Visiting Scientist, National Solar Observatory, 3665 Discovery Dr., Boulder, CO 80303, USA
     \and	
   Universidade Federal do Rio Grande do Norte, UFRN, Dep. de F\'{\i}sica, DFTE, CP1641, 59072-970, Natal, RN, Brazil 
    \and	
	Harvard-Smithsonian Center for Astrophysics, Cambridge, Massachusetts 02138, USA
          	   \and
   Stellar Astrophysics Centre, Department of Physics and Astronomy, Aarhus University, Ny Munkegade 120, 8000, Aarhus C, Denmark
                  }         
    	
   \date{Received XX XX XX; accepted XX XX XX}

% \abstract{}{}{}{}{} 
% 5 {} token are mandatory
 
  \abstract  
  {We identify a set of 18 solar analogs among the seismic sample of solar-like stars observed by the {\it Kepler} satellite rotating between 10 and 40~days. This set is constructed using the asteroseismic stellar properties derived using either the global oscillation properties or the individual acoustic frequencies.
  We measure the magnetic activity properties of these stars using observations collected by the photometric {\it Kepler} satellite and  by the ground-based, high-resolution \textsc{Hermes} spectrograph mounted on the \textsc{mercator} telescope.
The photospheric ($S_\mathrm{ph}$) and chromospheric (\Ssymbol~index) magnetic activity levels of these seismic solar analogs  are estimated and compared in relation to the solar activity. 
We show that the activity of the Sun is comparable to the activity of the seismic solar analogs, within the maximum-to-minimum temporal variations of the 11-year solar activity cycle~23.
In agreement with previous studies, the youngest stars and fastest rotators in our sample are actually the most active. The activity of stars older than the Sun seems to not evolve much with age.
Furthermore, the comparison of the photospheric, $S_\mathrm{ph}$, with the well-established chromospheric, \Ssymbol~index, indicates that the $S_\mathrm{ph}$ index can be used to provide a suitable magnetic activity proxy which can be easily estimated for a large number of stars from space photometric observations.
}

   \keywords{stars: solar-type -- stars: activity -- stars: evolution  -- methods: data analysis -- methods: observational}

   \maketitle
%
%________________________________________________________________

% -----------------------------------------------------------------------------------------
\section{Introduction}
 The Sun is a G-type star with a quasi-periodic 11-year magnetic
activity cycle that is most apparent in the rise and fall of the
occurrence of small regions of strong magnetic field which appear as
dark spots on the surface in the visible wavelengths \citep[][and
  references therein]{hathaway15}.  This behavior is believed to be
the result of an internal dynamo process driven by the rotational and
convective motions of the plasma in the outer convective layers of the
star \citep[][and references therein]{charb10}.  Numerous
physical models have been proposed to govern the amplification and
reconfiguration of global-scale toroidal and poloidal field
components in the observed cyclic fashion, yet as of today no \emph{ab
  initio} physical model exists which can explain all of the variety
of behavior in the sunspot cycle, and parameterized/empirical
models do not yet reliably predict basic features such as cycle
amplitudes or durations \citep[][and references therein]{petrovay10}.  

The question of whether variable magnetic activity occurs in other
stars was answered in the affirmative using long-term synoptic
observations of the \ion{Ca}{II} H and K lines. 
Over decades, extensive observational projects at Mount Wilson Observatory \citep[MWO;][]{wilson78,duncan91} and at Lowell Observatory \citep{hall07} were dedicated to measure the strength and the modulation of plasma emissions in the stellar chromosphere. These emissions, resulting from the non-thermal
heating that occurs in the presence of strong magnetic fields, are quantified through the so-called \Ssymbol \citep{wilson78} and $R'_{\rm HK}$ \citep{noyes84} indices. They strongly correlate with the presence of solar plage areas
surrounding active regions, and thereby make for excellent proxies of magnetic activity \citep{keil84}.

With stellar data on magnetic activity, it was then possible to study
the operation of dynamos in objects with different fundamental
properties, such as mass and rotation rate.  The MWO \ion{Ca}{II} H and K data were used
to explore the relationship between magnetic activity and rotation,
and it was found that stars that rotate faster have higher
magnetic activity \citep{vaughan81,baliunas83,noyes84}. In addition, it was observed that activity decreases with stellar ages \citep{sku72,soderblom91}.
Activity is also sensitive to stellar mass. Indeed, for stars with similar
rotation the less massive stars tend to have higher activity
\citep{noyes84}. \citet{baliunas95} reported the results of 25
years of MWO observations for a set of 111 lower main-sequence stars
which revealed a range of long-term variable behavior, from solar-like
and shorter cycles, to flat-activity and irregular variables.
\citet{saar92} and \citet{soon93} reported two distinct branches
of cycling stars, the ``active'' and ``inactive'' branches with 
respectively a higher and lower number of rotations per cycle.
\citet{brand98} and \citet{saar99} studied these dynamo
branches with a larger sample of stars in terms of dimensionless
quantities relevant to mean-field dynamo theory, and
\citet{bohm07} observed that the Sun lies squarely between
the two branches in a high-quality subset of the same sample, thus
appearing as a peculiar outlier among the cycling stars.  
In addition, \citet{lockwood07} compared the photospheric variability in
the visible Str{\"o}mgren $b$ and $y$ bands to the magnetic activity, and
found the Sun to be near the border in activity by which stars are
either ``spot-dominated'' and have visible emission anti-correlated
with magnetic activity, and those which are ``faculae-dominated'' and
have a positive correlation.
Whether the solar dynamo and the related surface magnetic activity are typical or
peculiar remains an open question which can only be answered by
the careful study of solar-analog stars.  The answer to this question
is relevant to dynamo theory, which may have more difficulty
explaining the behavior of an outlier. 

Finding solar-analog stars with fundamental properties as close as possible to the Sun \citep{cayrel96} and studying the characteristics of their surface magnetic activity is a very promising way to understand the solar variability and its associated dynamo processes. 
Moreover, a more detailed knowledge of the magnetic activity of solar analogs is also important for understanding the evolution of the Sun and its environment in relation to other stars and the habitability of planets.
Nevertheless, the identification of solar-analog stars depends on the accuracy of the estimated fundamental stellar parameters. 
The unprecedented quality of the continuous 4-year photometric observations collected by the {\it Kepler} satellite \citep{borucki10} allowed the measurements of acoustic oscillations in hundreds of solar-like stars \citep{chaplin14}. Indeed, the addition of asteroseismic data
was proven to provide the most accurate fundamental properties that can be derived from stellar modeling today, either from global oscillation properties \citep{chaplin14} or from fitted individual frequencies \citep{mathur12,metcalfe14}, compared to empirical scaling-law relations \citep{kjel95}. In the near future, even tighter constraints on stellar parameters will be met with the use of the astrometric observations collected by the GAIA satellite \citep{perryman01}.

Furthermore, the knowledge of the stellar rotation period is a key parameter in order to explain stellar activity \citep[][and references therein]{brun15}. The passage of spots of different sizes at different latitudes as the star rotates induces a modulation in its luminosity which is used to infer the rotation period of the stars. The measurements of thousands of stellar rotation periods of main-sequence stars \citep{nielsen13,reinhold13,mcq13,mcq14} were obtained from the {\it Kepler} observations.
Accurate periods of surface rotation were also measured for 310 \citep{garcia14} of the 540 pulsating solar-like field {\it Kepler} stars \citep{chaplin14} and for 11 \citep{ceillier16} of the 27 exoplanet-host stars ({\it Kepler} Objects of Interest, KOI) modeled by \citet{victor15} using asteroseismic measurements. A detailed comparison between the different methods developed to extract the surface rotation can be found in  \citet{aigrain15}. 

Based on the rotation period of the star, \citet{mathur14a} defined a photospheric proxy of stellar magnetic variability, called $S_\mathrm{ph}$, derived from the analysis of the light-curve fluctuations. Its calculation was adapted from the starspot proxy proposed by \citet{garcia10} where they showed that the photometric proxy of the F-type HD~49933 observed by the Convection, Rotation, and planetary Transits \citep[CoRoT;][]{baglin06} satellite is correlated with its magnetic activity. 
As the variability in the light curves can have different origins and timescales (magnetic activity, convection, oscillations, companion), the rotation period needs to be taken into account when calculating a photometric magnetic proxy.  
The photospheric surface proxy $S_\mathrm{ph}$ can be thus used to estimate the magnetic activity of the {\it Kepler} \citep{mathur14b,garcia14,salabert16} and CoRoT \citep{ferreira15} targets, where, for instance, spectral observations used to derive established activity proxies (as the \Ssymbol and $R'_{\rm HK}$ indices) do not exist and are difficult to obtain for a large number of faint stars.
Other photospheric metrics were developed to study the stellar variability in the {\it Kepler} data but they do not use the knowledge of the rotation rate in their definition \citep{basri10,basri11,chaplin11,campante14}.

In addition, Salabert et al. (in prep.) measured the $S_\mathrm{ph}$ index for the Sun using observations from the Variability of Solar Irrandiance and Gravity Oscillations  \citep[VIRGO;][]{frohlich95} instrument onboard the {\it Solar and Heliospheric Observatory} \citep[SoHO;][]{domingo95} spacecraft.
The VIRGO instrument is composed of three Sun photometers (SPM) at 402\,nm (the {\sc blue} channel), 500\,nm (the {\sc green} channel), and 862\,nm (the {\sc red} channel).
The activity index of the Sun thus derived was shown to be very well correlated with common solar surface activity proxies, such as, among others, the sunspot number, the 10.7-cm radio flux,  and the chromospheric Ca K line emission. 
Such photospheric proxy can be thus used to compare the magnetic activity of the Sun to solar-analog stars observed by {\it Kepler}.

In this work, we identify a set of solar analogs as defined in Section~\ref{sec:def} and observed by the {\it Kepler} satellite among the seismic sample of solar-like stars from \citet{chaplin14}. This set is constructed using the asteroseismic stellar properties found in the literature and derived using either the global oscillation properties \citep{chaplin14} and the individual acoustic frequencies \citep{mathur12, metcalfe14}, and for which a surface rotation period was measured \citep{garcia14}. In Section~\ref{sec:data}, we describe the set of observations used in this analysis to measure the magnetic activity properties of these stars, i.e. observations collected by the photometric {\it Kepler} satellite and by the spectroscopic \Hermes instrument \citep{raskin11,raskin11phd}. In Sections~\ref{sec:sph} and \ref{sec:chromo}, we respectively estimate the photospheric ($S_\mathrm{ph}$) and chromospheric (\Ssymbol~index) magnetic activity levels of these seismic solar analogs, and compare them in relation to the observed solar magnetic activity. We discuss the results in Section~\ref{sec:discussion} and conclude in Section~\ref{sec:conclusion}.
% ----------------------------------------------------------------------------------------- 

% -----------------------------------------------------------------------------------------
\begin{table*}[ht]
\begin{minipage}{\textwidth}
\caption{References of the initial {\it Kepler} sample of seismic solar-like stars with measured stellar rotation and the identified seismic solar analogs.}
\label{table:source}      
\centering               
\renewcommand{\footnoterule}{}  % to avoid a line before footnotes               
\begin{tabular}{l c c c}
\hline\hline    
 References &  Initial number of & Stars with & Number of identified \\
&  seismic solar-like stars & a measured $P_\mathrm{rot}$\footnote{The surface stellar rotation periods, $P_\mathrm{rot}$, were measured by \citet{garcia14}.}  & seismic solar analogs  \\
\hline
\citet{chaplin14}\footnote{The stellar parameters were derived using the global oscillation properties extracted from the one-month {\it Kepler} survey mode.} & &  &\\
\multicolumn{1}{l}{$-$ with complementary photometric data\footnote{The photometric observations were obtained through the SDSS survey \citep{pins12}.}} & 503 & 310 & 11 \\
\multicolumn{1}{l}{$-$ with complementary spectroscopic data\footnote{The spectroscopic observations were collected with the ESPaDOnS and NARVAL instruments \citep{bruntt12}.}} & 87 & 57 & 2\\
\hline
\citet{mathur12}\footnote{The stellar parameters were derived using the individual seismic frequencies extracted from the one-month {\it Kepler} survey mode and spectroscopic analysis.} & 22 & 12 & 2 \\
\hline
\citet{metcalfe14}\footnote{The stellar parameters were derived using the individual seismic frequencies extracted from nine-month {\it Kepler} time series by \citet{app12} and spectroscopic analysis \citep{bruntt12}.}  & 42 & 32 &  3\\
\hline             
\end{tabular}
\end{minipage}
\end{table*}
% -----------------------------------------------------------------------------------------

% ------------TABLE STELLAR PROPERTIES ------------------------------------
% Include table stellar properties
\begin{table*}
\caption{Seismic solar analogs observed by {\it Kepler} and their derived stellar properties.}
\label{table:prop}
\centering
\begin{tabular}{c c c c c c c c c c}
\hline \hline
$\#$ star & KIC & $M$ ($M_\sun$) & $R$ ($R_\sun$) & Age (Gyr) & $T_\text{eff}$ (K)  & log $g$ & $\Delta\nu$ ($\mu$Hz) & Ref. & {\bf Sources}\\
\hline
1 & 3241581 & 0.89 $\pm$ 0.12 & 1.03 $\pm$ 0.05 & 10.5 $\pm$ 5.0 & 5770 $\pm$ 81 & 4.36 $\pm$ 0.02 & 122.9 $\pm$ 1.6 & 1 & 1 \\
2 & 3656476 & 1.09 $\pm$ 0.01 & 1.32 $\pm$ 0.03 & 7.7 $\pm$ 0.2 & 5710 $\pm$ 84 & 4.22 $\pm$ 0.01 & 93.3 $\pm$ 1.3 & 3 & 1,2,3 \\
3 & 4914923 & 1.10 $\pm$ 0.01 & 1.37 $\pm$ 0.05 & 6.2 $\pm$ 0.2 & 5905 $\pm$ 84 & 4.21 $\pm$ 0.01 & 88.6 $\pm$ 0.3 & 3 & 3 \\
4 & 5084157 & 1.06 $\pm$ 0.13 & 1.36 $\pm$ 0.08 & 7.8 $\pm$ 3.4 & 5803 $\pm$ 83 & 4.19 $\pm$ 0.02 & 87.3 $\pm$ 2.6 & 1 & 1 \\
5 & 5774694 & 1.06 $\pm$ 0.05 & 1.00 $\pm$ 0.03 & 1.9 $\pm$ 1.8 & 5875 $\pm$ 84 & 4.46 $\pm$ 0.02 & 140.2 $\pm$ 4.0 & 2 & 2 \\
6 & 6116048 & 1.01 $\pm$ 0.03 & 1.22 $\pm$ 0.01 & 6.2 $\pm$ 0.4 & 5935 $\pm$ 84 & 4.27 $\pm$ 0.01 & 100.9 $\pm$ 1.4 & 4 & 2,3,4 \\
7 & 6593461 & 0.94 $\pm$ 0.16 & 1.29 $\pm$ 0.07 & 10.7 $\pm$ 4.4 & 5817 $\pm$ 101 & 4.19 $\pm$ 0.02 & 90.8 $\pm$ 2.0 & 1 & 1 \\
8$^\dagger$ & 7296438 & 0.91 $\pm$ 0.15 & 1.30 $\pm$ 0.06 & 12.3 $\pm$ 4.3 & 5749 $\pm$ 56 & 4.17 $\pm$ 0.03 & 88.6 $\pm$ 2.1 & 1 & 1 \\
9 & 7680114 & 1.12 $\pm$ 0.07 & 1.43 $\pm$ 0.04 & 6.5 $\pm$ 1.5 & 5855 $\pm$ 84 & 4.18 $\pm$ 0.01 & 85.1 $\pm$ 1.3 & 2 & 1,2 \\
10 & 7700968 & 1.00 $\pm$ 0.12 & 1.21 $\pm$ 0.06 & 7.5 $\pm$ 3.1 & 5982 $\pm$ 75 & 4.27 $\pm$ 0.02 & 102.6 $\pm$ 2.9 & 1 & 1 \\
11 & 9049593 & 1.13 $\pm$ 0.14 & 1.40 $\pm$ 0.06 & 6.4 $\pm$ 3.4 & 5788 $\pm$ 59 & 4.19 $\pm$ 0.02 & 86.3 $\pm$ 2.1 & 1 & 1 \\
12 & 9098294 & 1.00 $\pm$ 0.03 & 1.15 $\pm$ 0.01 & 7.3 $\pm$ 0.5 & 5840 $\pm$ 84 & 4.30 $\pm$ 0.01 & 108.8 $\pm$ 1.7 & 4 & 1,2,4 \\
13 & 10130724 & 0.85 $\pm$ 0.12 & 1.08 $\pm$ 0.05 & 13.8 $\pm$ 5.0 & 5648 $\pm$ 70 & 4.31 $\pm$ 0.02 & 112.9 $\pm$ 1.6 & 1 & 1 \\
14 & 10215584 & 0.99 $\pm$ 0.13 & 1.12 $\pm$ 0.05 & 6.8 $\pm$ 3.5 & 5995 $\pm$ 57 & 4.34 $\pm$ 0.02 & 114.7 $\pm$ 3.1 & 1 & 1 \\
15 & 10644253 & 1.13 $\pm$ 0.05 & 1.11 $\pm$ 0.02 & 1.1 $\pm$ 0.2 & 6030 $\pm$ 84 & 4.40 $\pm$ 0.01 & 123.6 $\pm$ 2.7 & 4 & 2,4 \\
16 & 10971974 & 1.04 $\pm$ 0.12 & 1.19 $\pm$ 0.06 & 5.8 $\pm$ 3.0 & 6030 $\pm$ 58 & 4.30 $\pm$ 0.02 & 106.2 $\pm$ 3.5 & 1 & 1 \\
17$^\dagger$ & 11127479 & 1.14 $\pm$ 0.12 & 1.36 $\pm$ 0.06 & 5.1 $\pm$ 2.2 & 5998 $\pm$ 57 & 4.22 $\pm$ 0.01 & 90.6 $\pm$ 2.5 & 1 & 1 \\
18 & 11971746 & 1.11 $\pm$ 0.14 & 1.35 $\pm$ 0.06 & 6.0 $\pm$ 2.8 & 5952 $\pm$ 75 & 4.22 $\pm$ 0.02 & 90.8 $\pm$ 2.1 & 1 & 1 \\
\hline
\end{tabular}
\tablebib{(1) global oscillation properties from one-month data and photometry \citep{chaplin14}; (2) global oscillation properties from one-month data and spectroscopy \citep{chaplin14}; (3) individual acoustic frequencies from one-month data and spectroscopy \citep{mathur12} ; (4) individual acoustic frequencies from nine-month data\footnote{From \citet{app12}} and spectroscopy \citep{metcalfe14}. The large frequency separations $\Delta\nu$ were taken from \citet{chaplin14}.}
\tablefoot{The two stars indicated by the symbol $^\dagger$ are candidates for hosting a planetary system from the {\it Kepler} Objects of Interest (KOI) list (source: MAST at \url{https://archive.stsci.edu/kepler/}). From the analysis of spectroscopic \Hermes observations, \citet{beck15} found for KIC\,3241581 the following values: $T_\text{eff}=5769\pm4$\,K, log\,$g=4.39\pm0.01$, $M=1.03\pm0.10M_\sun$, and $R=1.08\pm0.10R_\sun$, and \citet{salabert16} found for KIC\,10644253: $T_\text{eff}=6006\pm100$\,K, and log\,$g=4.3\pm0.1$.}
\end{table*}

% -----------------------------------------------------------------------------------------

% -----------------------------------------------------------------------------------------
\section{Defining the sample of seismic solar analogs}
\label{sec:def}
\citet{cayrel96} provided a definition of a solar analog based on the fundamental parameters of the star such as the mass and the effective temperature. However, any definition depends on the accuracy of the measured properties and their intrinsic importance in the characterization of a solar analog. Today, asteroseismology combined with high-resolution spectroscopy has allowed to substantially improve the accuracy of the stellar parameters and to reduce their errors \citep{mathur12,chaplin14,metcalfe14}. Furthermore, \citet{chaplin11} showed evidences that the magnetic activity inhibits the amplitudes of solar-like oscillations. Indeed, in the case of the Sun, the amplitudes of the acoustic modes decrease by about 15\% between solar minimum and maximum for the low-degree modes \citep[see e.g.,][and references therein]{salabert03}, which are the modes detectable in observations of solar-like stars. In consequence, the selection of stars analog to the Sun should include the detection of solar-like oscillations as an additional selection criterium. This is what we called a seismic solar-analog star, thus extending the \citet{cayrel96}'s definition.

In this work, we also took into account the typical observational asteroseismic and photometric/spectroscopic uncertainties returned on the derived stellar parameters. 
The mass, $M$, has indeed to be within $\pm$\,10\% of the solar mass to which we added 5\% which corresponds to the mean errors on the seismic masses derived using individual oscillation frequencies \citep{metcalfe14}. The effective temperature, $T_\mathrm{eff}$, has to be within $\pm$\,150\,K to the solar effective temperature, to which we account for a typical error of 100\,K \citep{bruntt12}. The surface gravity, log $g$, must be within $\pm$\,0.3\,dex of the solar value. The selection was thus based on the following: 
(1) $0.85M_\sun\,\leq\,M\,\leq\,1.15M_\sun$; (2) 5520\,K\,$\leq\,T_\mathrm{eff}\,\leq$\,6030\,K; and (3) 4.14\,$\leq\,\mathrm{log}\,g\,\leq$\,4.74. Moreover, we included in the sample only the stars with a measured rotation period \citep{garcia14} to ensure the presence of magnetic activity. In addition, we limited also the selection to stars with a large frequency separation $\Delta\nu\,\geq\,85\,\mu$Hz in order to avoid evolved stars. For comparison, $\Delta\nu_\sun\,\simeq\,135$\,$\mu$Hz for the Sun \citep[e.g.,][]{grec80}. The values of $\Delta\nu$ were taken from \citet{chaplin14} which were estimated from one month each of the short-cadence (SC) {\it Kepler} survey mode.
Nonetheless, we checked that consistent values within the error bars were obtained when analyzing the entire SC data collected over the $\sim$\,4 years of the  {\it Kepler}  mission using the A2Z pipeline \citep{mathur10}. We note also that the solar values were taken to be: $T_\mathrm{eff}$\,=\,5777\,K and log\,$g$\,=\,4.44.

A total of 18 seismic solar analogs were then identified from the photometric {\it Kepler} observations of the solar-like oscillators \citep{chaplin14}. To do so, we chose the stellar parameters estimated using the individual frequencies whenever available, because they were derived with a better precision \citep{metcalfe14}. Otherwise we used the results obtained from the analysis of the global oscillation properties. 
Among these 18 stars, the stellar parameters of 5 of them were derived with the Asteroseismic Modeling Portal \citep[AMP;][]{metcalfe09} 
using individual frequencies and spectroscopy \citep{mathur12,metcalfe14}.
The remaining 13 seismic solar analogs were identified through the stellar parameters derived using grid modeling of the global oscillation properties \citep{chaplin14}, for 11 of them combined with the atmospheric properties from the SDSS photometric survey \citep{pins12}, and for 2 of them with the spectroscopic properties obtained with ESPaDOnS and NARVAL observations \citep{bruntt12}. However, the accuracy on the associated stellar parameters derived with photometry is less than for the ones estimated using spectroscopic analysis.
 
Table~\ref{table:source} summarizes the different literature sources of the stellar parameters of the oscillating solar-like stars used in the initial sample and the associated numbers of identified seismic solar analogs with measured surface rotation, while their corresponding KIC numbers and stellar properties are given in Table~\ref{table:prop}. The references of the derived stellar parameters used in this work are provided by the column "Ref.": 1 when global oscillation properties and photometry were used \citep{chaplin14}; 2 when global oscillation properties and spectroscopy were used \citep{chaplin14}; 3 and 4  when individual acoustic frequencies and spectroscopy were used, respectively in \citet{mathur12} and \citet{metcalfe14}.  
Finally, some seismic solar analogs were identified in more than one literature source, as indicated in the column "Sources". In that case, we favored the stellar parameters estimated with a known better accuracy, whose reference is given in the column "Ref.".
We note also that all the 18 seismic solar analogs identified in this work from the {\it Kepler} sample rotate between 10 and 40\,days, with only 7 stars rotating faster than 20\,days. Finally, the crowding factor of the selected stars was checked to be less than 1\%, which implies almost no risk of pollution by nearby field stars. 
 
We also note that two of these stars were studied with greater details. A comprehensive differential analysis of \Hermes spectroscopic observations of KIC\,3241581, with seismic properties similar to the Sun, was performed by \citet{beck15}, revealing that this is actually a binary system. Both the photospheric activity and the p-mode oscillation frequencies of the young (1\,Gyr-old) solar analog KIC\,10644253 were shown to vary with a temporal modulation of about 1.5 years \citep{salabert16}. In addition, two stars, KIC\,7296438 and KIC\,11127479, are reported as candidates for hosting a planetary system in the KOI list.
% -----------------------------------------------------------------------------------------

% -----------------------------------------------------------------------------------------
\section{Observations}
\label{sec:data}
\subsection{Photometric observations from the {\it Kepler} satellite}
\label{sec:photokepler}
The long-cadence (LC) observations of temporal sampling of 29.4244\,min collected by the \emph{Kepler} satellite over almost the entire duration of the mission from 2009 June 20 (Quarter\,2, hereafter Q2) to 2013 May 11 (Q17) (i.e., a total of 1422 days) were used in this analysis. The two first quarters Q0 ($\sim$\,10 days) and Q1 ($\sim$\,33 days) were dropped because of calibration problems since it is performed quarter by quarter. All the light curves were corrected for instrumental problems using the \emph{Kepler} Asteroseismic Data Analysis and Calibration Software \citep[KADACS,][]{garcia11}. The time series were then high-pass filtered in a quarter by quarter basis using a triangular smooth function with a cut off of 55\,days (see Appendix~\ref{app:1} regarding the choice of the filter). The signature of photospheric magnetic activity can be thus inferred by studying the temporal fluctuations of the light curves.

% -----------------------------------------------------------------------------------------
\subsection{Spectroscopic observations from the  \Hermes instrument}
\label{sec:hermes_obs}
Complementary high-resolution spectroscopic observations (R\,=\,85\,000) of the 18 identified seismic solar analogs were obtained with the \Hermes spectrograph, mounted to the 1.2-m \textsc{Mercator} telescope at the Observatorio del Roque de los Muchachos (La Palma, Canary Islands, Spain). For 16 of the stars, the \Hermes observations were performed over two observing runs of three days each in June and July 2015, covering a temporal interval, $\Delta$T, of about 35\,days. The 2 other targets,  KIC\,3241581 and KIC\,10644253, were put on a long-term monitoring program with the \Hermes spectrograph and thus longer temporal coverages are available \citep{beck15,salabert16}. Observations between April 2014 and September 2015 ($\Delta$T\,=\,504~days) and between March 2015 and September 2015 ($\Delta$T\,=\,180~days) were respectively used in this analysis. Details are provided in Table~\ref{table:sph}. The data reduction of the raw observations was performed with the instrument-specific pipeline \citep{raskin11}. The covered wavelength range is between 375 and 900\,nm.
The radial velocity, RV,  was measured for each individual spectrum with a cross-correlation analysis using the G2-mask in the \Hermes pipeline toolbox. See \citet{beck15} for more details regarding the data processing of the \Hermes observations. 

% ------ TABLE ACTIVITY -------
% Include table magnetic activity results
\begin{table*}
\caption{Photospheric and chromospheric activity properties of the 18 seismic solar analogs derived from the photometric {\it Kepler} and spectroscopic \Hermes observations.}
\label{table:sph}
\centering
\begin{tabular}{c c c c c c c c c c c c c c}
\hline \hline
$\#$ star & KIC & $P_\mathrm{rot}$ (days) & $S_\mathrm{ph}$ (ppm) & $\mathcal{S}$ index & log $R^\prime_\mathrm{HK}$ & SNR(Ca) & $\Delta$T (days) & N$_\text{spec}$\\
\hline
1 & 3241581 & 26.3 $\pm$ 2.0 & 236.9 $\pm$ 3.8 & 0.182 $\pm$ 0.024 & $-4.868 \pm 0.164$ & 17.1 & 504 & 23 \\
2 & 3656476 & 31.7 $\pm$ 3.5 & 89.1 $\pm$ 1.6 & 0.168 & $-5.002 \pm 0.100$ & 24.5 & <1 & 1 \\
3 & 4914923 & 20.5 $\pm$ 2.8 & 139.1 $\pm$ 2.5 & 0.168 $\pm$ 0.014 & $-4.918 \pm 0.100$ & 26.6 & 34 & 6 \\
4 & 5084157 & 22.2 $\pm$ 2.8 & 270.8 $\pm$ 4.8 & 0.217 $\pm$ 0.107 & $-4.840 \pm 0.580$ & 11.0 & 33 & 9 \\
5 & 5774694 & 12.1 $\pm$ 1.0 & 2492.3 $\pm$ 61.2 & 0.269 $\pm$ 0.021 & $-4.581 \pm 0.059$ & 23.7 & 36 & 3 \\
6 & 6116048 & 17.3 $\pm$ 2.0 & 90.2 $\pm$ 1.9 & 0.157 $\pm$ 0.001 & $-4.984 \pm 0.042$ & 24.1 & 35 & 3 \\
7 & 6593461 & 25.7 $\pm$ 3.0 & 188.8 $\pm$ 3.6 & 0.260 $\pm$ 0.119 & $-4.550 \pm 0.333$ & 10.4 & 35 & 8 \\
8$^\dagger$ & 7296438 & 25.2 $\pm$ 2.8 & 177.1 $\pm$ 3.0 & 0.162 & $-4.994 \pm 0.151$ & 20.4 & 35 & 3$^a$ \\
9 & 7680114 & 26.3 $\pm$ 1.9 & 50.2 $\pm$ 1.0 & 0.185 $\pm$ 0.019 & $-4.881 \pm 0.164$ & 19.4 & 35 & 3 \\
10 & 7700968 & 36.2 $\pm$ 4.2 & 88.9 $\pm$ 1.4 & 0.181 $\pm$ 0.010 & $-4.866 \pm 0.123$ & 19.8 & 35 & 3 \\
11 & 9049593 & 12.4 $\pm$ 2.5 & 290.8 $\pm$ 6.4 & 0.177 $\pm$ 0.016 & $-4.955 \pm 0.200$ & 18.4 & 35 & 3 \\
12 & 9098294 & 19.8 $\pm$ 1.3 & 292.7 $\pm$ 5.2 & 0.203 $\pm$ 0.031 & $-4.737 \pm 0.136$ & 19.5 & 35 & 3 \\
13 & 10130724 & 32.6 $\pm$ 3.0 & 155.0 $\pm$ 2.5 & 0.252 $\pm$ 0.067 & $-4.548 \pm 0.246$ & 10.6 & 34 & 5 \\
14 & 10215584 & 22.2 $\pm$ 2.9 & 138.3 $\pm$ 2.5 & 0.215 $\pm$ 0.049 & $-4.738 \pm 0.211$ & 17.4 & 35 & 4 \\
15 & 10644253 & 10.9 $\pm$ 0.9 & 549.4 $\pm$ 13.3 & 0.219 $\pm$ 0.014 & $-4.691 \pm 0.059$ & 30.1 & 180 & 12 \\
16 & 10971974 & 26.9 $\pm$ 4.0 & 309.9 $\pm$ 4.9 & 0.241 $\pm$ 0.066 & $-4.761 \pm 0.312$ & 10.0 & 35 & 3 \\
17$^\dagger$ & 11127479 & 17.6 $\pm$ 1.8 & 115.9 $\pm$ 2.8 & 0.271 $\pm$ 0.169 & $-4.683 \pm 0.482$ & 10.2 & 35 & 4 \\
18 & 11971746 & 19.5 $\pm$ 2.1 & 90.5 $\pm$ 2.0 & 0.152 $\pm$ 0.028 & $-4.937 \pm 0.217$ & 16.6 & 34 & 7 \\
\hline
\end{tabular}
\tablefoot{The surface rotation periods, $P_\mathrm{rot}$, were measured by \citet{garcia14} from the {\it Kepler} observations.  The columns $\Delta$T and N$_\text{spec}$ corresponds respectively to the interval of time covered by the \Hermes observations and the associated number of analyzed spectra. The two stars indicated by the symbol $^\dagger$ are candidates for hosting a planetary system from the {\it Kepler} Objects of Interest (KOI) list (source: MAST at \url{https://archive.stsci.edu/kepler/}). $^a$ In the case of KIC\,7296438, three \hermes spectra were collected, but only one spectrum has a S/N high enough to measure the \Ssymbol index.}
\end{table*}

% -------------------------

% ------ TABLE BINARIES -------
\begin{table}[t]
\caption{List of binary systems discovered with the \Hermes spectroscopic observations within the sample of seismic solar analogs.}
\label{table:rv}
\begin{center}
\begin{tabular}{c c c l}
\hline\hline
KIC & RV &$\Delta$RV &  Comment\\
 &  (km\,s$^{-1}$) &(km\,s$^{-1}$) &\\
\hline
3241581	&  $-30.68$ & $0.95$ & binary \citep{beck15}\\
4914923	& $-31.30$ & $1.05$ & candidate binary\\
7296438	& $-7.50$ & $16.57$ & binary \\
9098294	& $-43.19$ & $41.35$ 	& binary\\
\hline
\end{tabular}
\tablefoot{The mean radial velocity, RV, and the associated maximum difference, $\Delta$RV, are also given.}
\end{center}
\label{tab:binary}
\end{table}
% ------------------------------------------

Given the sparse sampling and the limited period of time, only stars with a maximum difference in radial velocity, $\Delta$RV, of more than 1\,km\,s$^{-1}$ are considered as binary candidates. Besides the confirmed binary KIC\,3241581 \citep{beck15}, three additional stars exceed this limit: KIC\,4914923, KIC\,7296438, and KIC\,9098294 (see Table~\ref{tab:binary}). The KIC\,7296438 and KIC\,9098294 systems are however the most noticeable with a $\Delta$RV of 16.57\,km\,s$^{-1}$ and of 41.35\,km\,s$^{-1}$ respectively. Moreover, none of these three systems shows eclipses or tidally induced flux modulation in their light curve. 
We also note that the system KIC\,7296438 is listed as a planetary host candidate (see Table~\ref{table:prop}). However, the detected variation in RV is only compatible with a stellar binary and not with a companion of substellar mass.
Moreover, the spectroscopic data suggest that there could be other long periodic binary systems in our sample but they are close to our current significance threshold. Therefore, more data are needed to verify the binary nature of those systems (Beck et al., in prep.). Nonetheless, we found that the magnetic activity of the detected binaries do not differ from the one derived for single field stars (see Table~\ref{table:sph}).
A detailed discussion of the observation of binary stars with solar-like oscillating primaries detected with the \Hermes  spectrograph can be found in \citet{beck14}. 
% -----------------------------------------------------------------------------------------

% -----------------------------------------------------------------------------------------
\section{Photospheric magnetic activity}
\label{sec:sph}
\subsection{ The $S_\mathrm{ph}$ proxy}
The photospheric activity proxy, $S_\mathrm{ph}$, is a measurement of stellar magnetic variability derived by means of the surface rotation, $P_\mathrm{rot}$ \citep{mathur14b,garcia14,ferreira15,salabert16}. We recall that the surface rotation can be inferred through the periodic luminosity fluctuations induced by the passage of spots in the line of sight \citep[see e.g.,][]{nielsen13,reinhold13,mcq13,mcq14,garcia14}.
The $S_\mathrm{ph}$ proxy is defined as the mean value of the light curve fluctuations estimated as the standard deviations calculated over subseries of length $5 \times P_\mathrm{rot}$. 
In this way, \citet{mathur14a} demonstrated that most of the measured variability is only related to the magnetism (i.e. the spots) and not to the other sources of variability at different timescales, such as convective motions, oscillations, stellar companion, or instrumental problems. However, we note that such proxy represents a lower limit of the stellar photospheric activity because it depends on the inclination angle of the rotation axis in relation to the line of sight. This is also assuming that the development of the latitudinal distribution of the starspots is comparable to the one observed for the Sun, i.e. from mid to low latitudes, which would underestimate the  $S_\mathrm{ph}$ for highly inclined stars. Finally, the error on $S_\mathrm{ph}$ was returned as the standard error of the mean value.

% -----------------------------------------------------------------------------------------
\begin{figure*}[tbp]
\begin{center} 
\includegraphics[width=0.49\textwidth,angle=0]{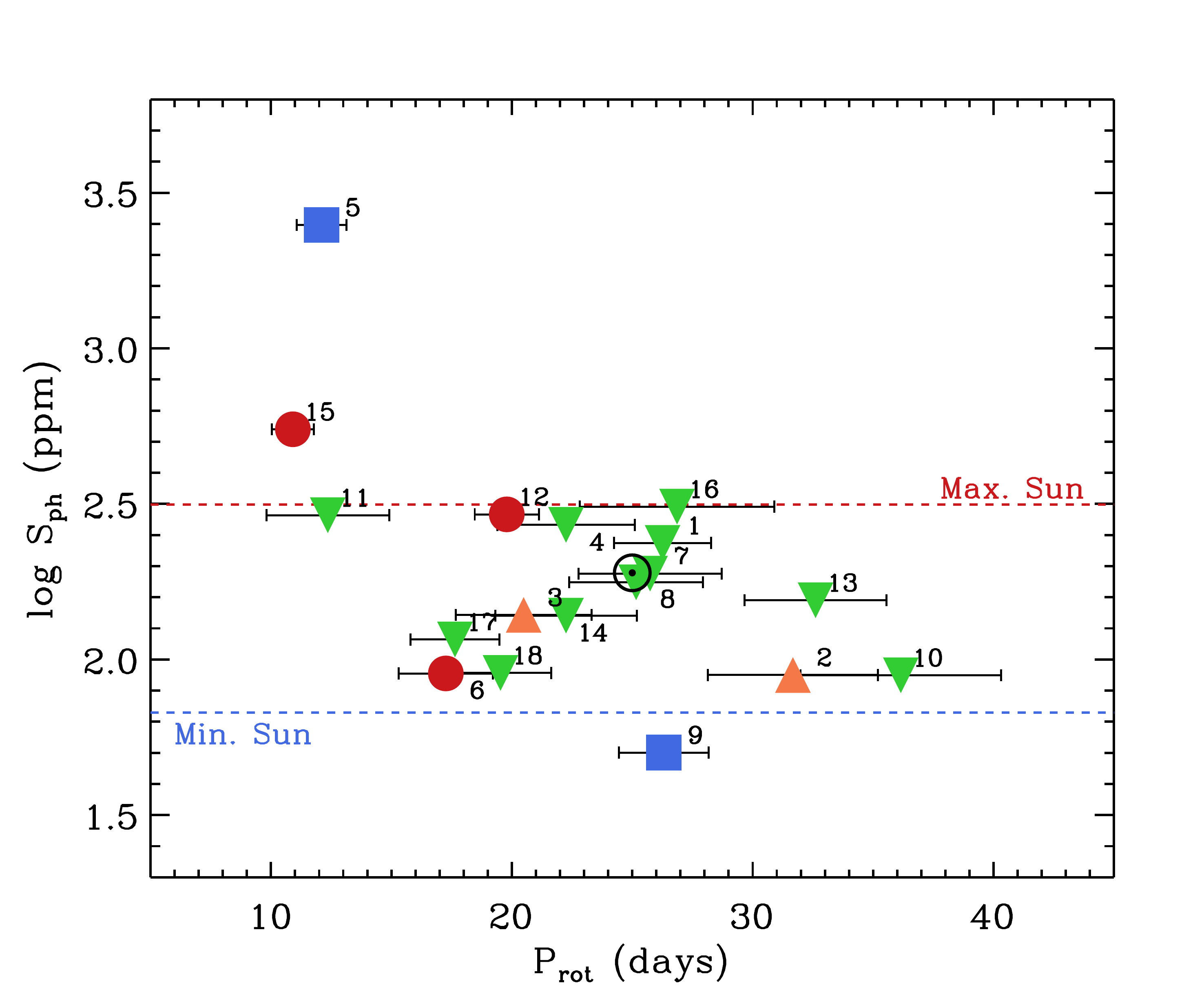}
\includegraphics[width=0.49\textwidth,angle=0]{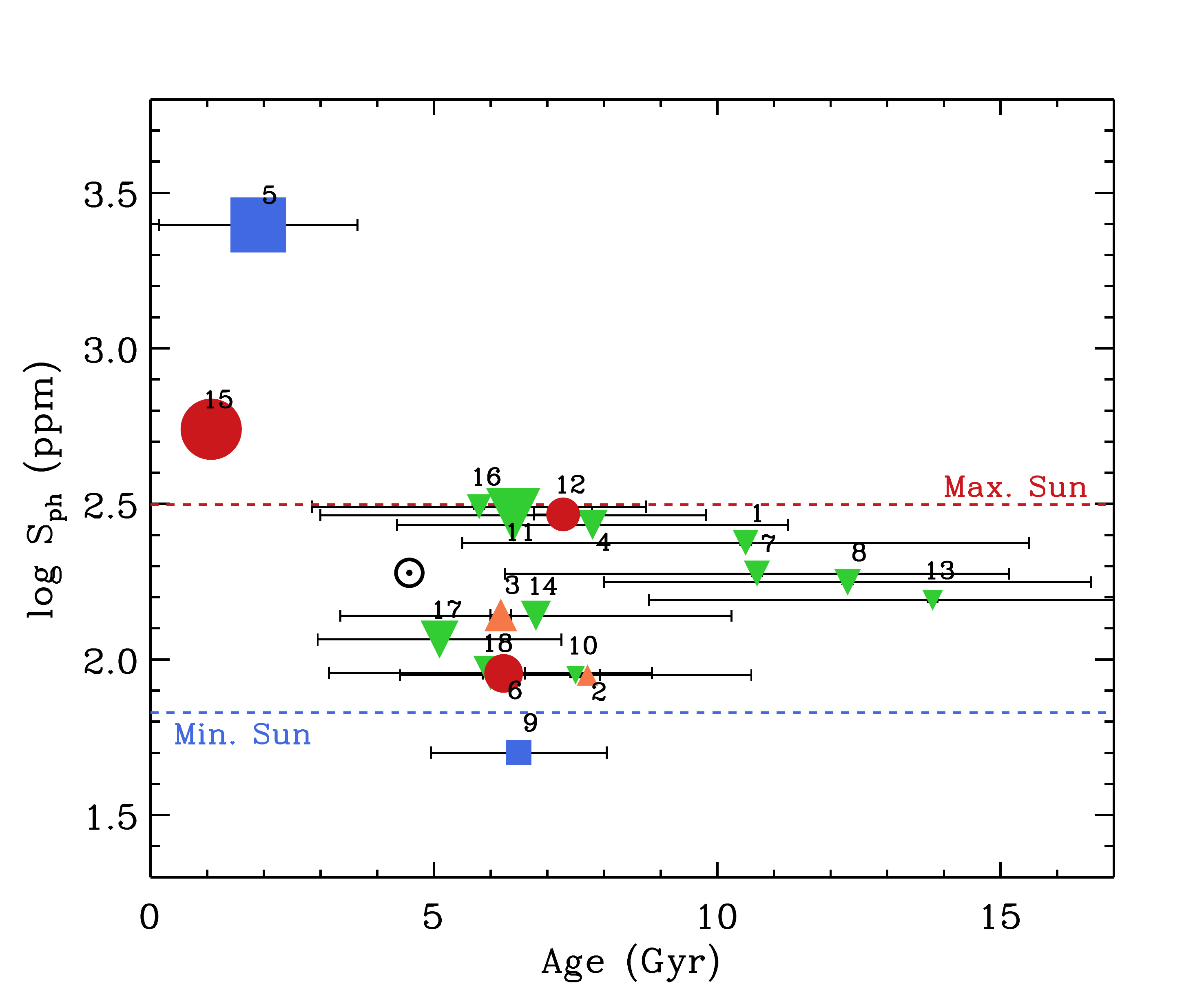}
\end{center}
\caption{\label{fig:kepler}  
(Left panel) Photospheric magnetic activity index, $S_{\text{ph}}$ (in ppm), as a function of the rotational period, $P_\mathrm{rot}$ (in days), of the 18 seismic solar analogs observed with the {\it Kepler} satellite. The mean activity level of the Sun calculated from the VIRGO/SPM observations is represented for a rotation of 25\,days with its astronomical symbol, and its mean activity levels at minimum and maximum of the 11-year cycle are represented by the horizontal dashed lines. (Right panel) Same as the left panel but as a function of the seismic age (in Gyr). The position of the Sun is also indicated for an age of 4.567\,Gyr. The size of the symbols is inversely proportional to the rotation period, $P_\mathrm{rot}$. In both panels, each star is referred by the same number as given in Tables~\ref{table:prop} and \ref{table:sph}.}
\end{figure*} 
% -----------------------------------------------------------------------------------------

% -----------------------------------------------------------------------------------------
\subsection{Relation with rotation and age}
The values of the photospheric activity proxy, $S_\mathrm{ph}$, of the 18 seismic solar analogs were then measured over 1422~days of {\it Kepler} observations. We used here the rotation periods, $P_\mathrm{rot}$, estimated by \citet{garcia14}.
The magnitude correction of the photon noise provided by \citet{jenkins10} was then applied to the $S_\mathrm{ph}$ for each of the analyzed star. The measured $S_\mathrm{ph}$ are given in Table~\ref{table:sph}, which summarizes the results of the magnetic activity properties of these stars. 
The left panel of Fig.~\ref{fig:kepler} shows on a y-log scale the photospheric magnetic activity levels, $S_\mathrm{ph}$, of the 18 seismic solar analogs as a function of their rotational periods, $P_\mathrm{rot}$.
The color and symbol codes correspond to the different sources of observations used to derive the stellar parameters, as described in Table~\ref{table:prop}: (1) green upside down triangles:  global oscillation properties and photometry \citep{chaplin14}; (2) blue squares: global oscillation properties and spectroscopy \citep{chaplin14}; (3) orange triangles: one-month individual acoustic frequencies and spectroscopy \citep{mathur12}; and (4) red dots: nine-month individual acoustic frequencies fitted by \citet{app12} and spectroscopy \citep{metcalfe14}. 

The mean value of the photospheric activity level of the Sun, $S_\mathrm{ph,\,Cycle\,23\,\sun}\,=\,189.9\,\pm\,0.5$\,ppm, calculated from the observations collected by the VIRGO/SPM photometers covering the entire cycle~23 (1996 -- 2008) for a equatorial rotation $P_\mathrm{rot,\,\sun}\,=\,25$\,days is also indicated (Salabert at al., in prep.). We only used data covering cycle~23, because during the unusually deep and extended activity minimum of cycle~24, the Sun reached activity values considerably lower than in any of its previously observed minima \citep[see][and references therein]{hathaway15}. In order to monitor the long-lived features on the solar surface, the VIRGO/SPM data were processed through the KADACS  pipeline as described in Section~\ref{sec:photokepler}. 
A composite photometric time series obtained by combining the observations from the {\sc green} and {\sc red} channels was used. Indeed, the {\it Kepler} broad bandpass includes these two channels at 500 and 862\,nm respectively \citep{basri10}.
In addition, the horizontal lines correspond to the photospheric magnetic activity levels of the Sun estimated at minimum and maximum of cycle~23, respectively $S_\mathrm{ph,\,\textsc{Min\,23}\,\sun}\,=\,67.4\,\pm\,0.2$\,ppm and $S_\mathrm{ph,\,\textsc{Max\,24}\,\sun}\,=\,314.5\,\pm\,0.8$\,ppm. These values were reported on Fig.~\ref{fig:kepler}.

The photospheric activity level of the identified seismic solar analogs falls for most of them (15 out of 18) within the range of activity covered between the minimum and the maximum of the solar cycle.
Two stars with a rotation period of about 10~days are more active than the Sun at its maximum: KIC\,5774694 and KIC\,10644253 with respective magnetic photospheric activity of $8 \times S_\mathrm{ph,\textsc{Max}\,\sun}$ and $2 \times S_\mathrm{ph,\textsc{Max}\,\sun}$.  Only one analog, KIC\,7680114, with a rotation period comparable to the Sun, is observed to have a photospheric activity slightly lower than the Sun at its minimum of the magnetic cycle by about $0.7 \times S_\mathrm{ph,\textsc{Min}\,\sun}$.

The right panel of Fig.~\ref{fig:kepler} shows for the same set of 18 stars the measured photospheric activity levels, $S_\mathrm{ph}$, as a function of their seismic ages \citep[taken from][and reported in Table~\ref{table:prop}]{mathur12,chaplin14,metcalfe14} compared to the Sun. The color code is the same as in the left panel of Fig.~\ref{fig:kepler} but the size of the symbols is inversely proportional to the surface rotation period, $P_\mathrm{rot}$. We note the large differences in the uncertainties of the age estimates depending on the applied method to derive them \citep[see][]{metcalfe14}. It is also worth noticing that stars between 2 and 5\,Gyr-old are missing in our sample.
Nevertheless, the position of the Sun indicates that its photospheric activity is compatible with older solar analogs. Although the {\it Kepler} seismic sample contains only few stars younger than the Sun \citep{mathur12,chaplin14,metcalfe14}, our results show that the two youngest seismic solar analogs in our sample below 2\,Gyr-old are actually the most active, as well as being the fastest rotating stars as mentioned above: KIC\,5774694 and KIC\,10644253. Furthermore, the photospheric activity of stars older than the Sun seems to not evolve much with age.
\citet{pace13} studied the chromospheric activity of field dwarf stars as an age indicator, and showed that it stops decaying for stars older than 2\,Gyr-old. 
The photospheric activity of the solar analogs analyzed in this work is observed to be comparable to the Sun after 2\,Gyr-old within the minimum-to-maximum range of the solar cycle, given the uncertainties on the age estimates and the limited size of our sample. This is also assuming that our sample is not too biased towards stars observed during periods of low activity of their magnetic cycles, or that they are not in an extended cycle. Nevertheless, observations of additional solar analogs younger than the Sun are needed to fill the range of ages below 5\,Gyr-old.

% -----------------------------------------------------------------------------------------
\subsection{Dependence on the inclination angle}
In the case of the Sun, spots are preferably formed between $\pm$\,55$\degr$ of latitude and appear closer and closer to the equator as the cycle progresses. This temporal and latitudinal development of sunspots is illustrated by the so-called butterfly diagram. In solar-like stars, preliminary observations of the spatial distribution of magnetic fields are accessible with spectropolarimetric data \citep[see e.g.,][and references therein]{alvarado15,folsom16}. Nevertheless, in photometric observations,  the signature of temporal magnetic fluctuations is dependent on the line of sight in relation to the inclination angle of the rotation axis of the star. 
The estimation of the inclination angle requires three independent observables: the projected rotational velocity, $\upsilon\,\mathrm{sin}\,i$, the rotation period, $P_\mathrm{rot}$, and the model-dependent radius, $R$. 
These three parameters are measured with intrinsic and different
accuracy with varying associated uncertainties. This thus makes difficult to have
accurate estimates of the inclination of the rotation axis  with satisfactory error bars in order to be able to calibrate the photospheric proxy, $S_\mathrm{ph}$, with the angle.
\citet{bruntt12} and \citet{molenda13} provided $\upsilon\,\mathrm{sin}\,i$ values measured from spectroscopic observations for some of our {\it Kepler} sample of seismic solar analogs, but unfortunately with large uncertainties. Asteroseismic estimates of $\upsilon\,\mathrm{sin}\,i$ were also obtained by \citet{doyle14} through the peak-fitting analysis of the oscillation modes for some of the stars studied here. 
Moreover, more accurate estimates of the asteroseismic $\upsilon\,\mathrm{sin}\,i$ will be derived for fast rotators as the rotational splittings between the mode components will be larger, while for slow rotators like the Sun, it is very challenging to determine \citep{ballot08}.

Most of the stars in our {\it Kepler} sample of solar analogs can be considered as slow rotators. Only three stars spin in less than 15~days. In the case of KIC\,10644253, the fastest rotator studied here ($P_\mathrm{rot}=10.9$~days), three independent values of its projected velocity were estimated from both spectroscopic \citep{bruntt12,salabert16} and asteroseismic \citep{doyle14} observations. They all suggest that this 1\,Gyr-old, young Sun is likely to be observed along a close pole-on angle. In consequence, the associated measured photospheric activity level, $S_\mathrm{ph}$, is most probably underestimated and should be considered as a lower limit.
For the second faster rotator ($P_\mathrm{rot}=12.1$~days) and most active star in our sample, KIC\,5774694, only one spectroscopic measurement from \citet{bruntt12} is available, which suggests that this star is observed close to its equatorial plan.
For the third one, KIC\,9049593 ($P_\mathrm{rot}=12.4$~days), no $\upsilon\,\mathrm{sin}\,i$ measurement was unfortunately found in the literature.

% -----------------------------------------------------------------------------------------
\section{Chromospheric magnetic activity}
\label{sec:chromo}

\subsection{Deriving the \Ssymbol~index from the \Hermes observations}
\label{sec:sindex}
\begin{figure}[t!]
\begin{center}
\includegraphics[width=0.47\textwidth]{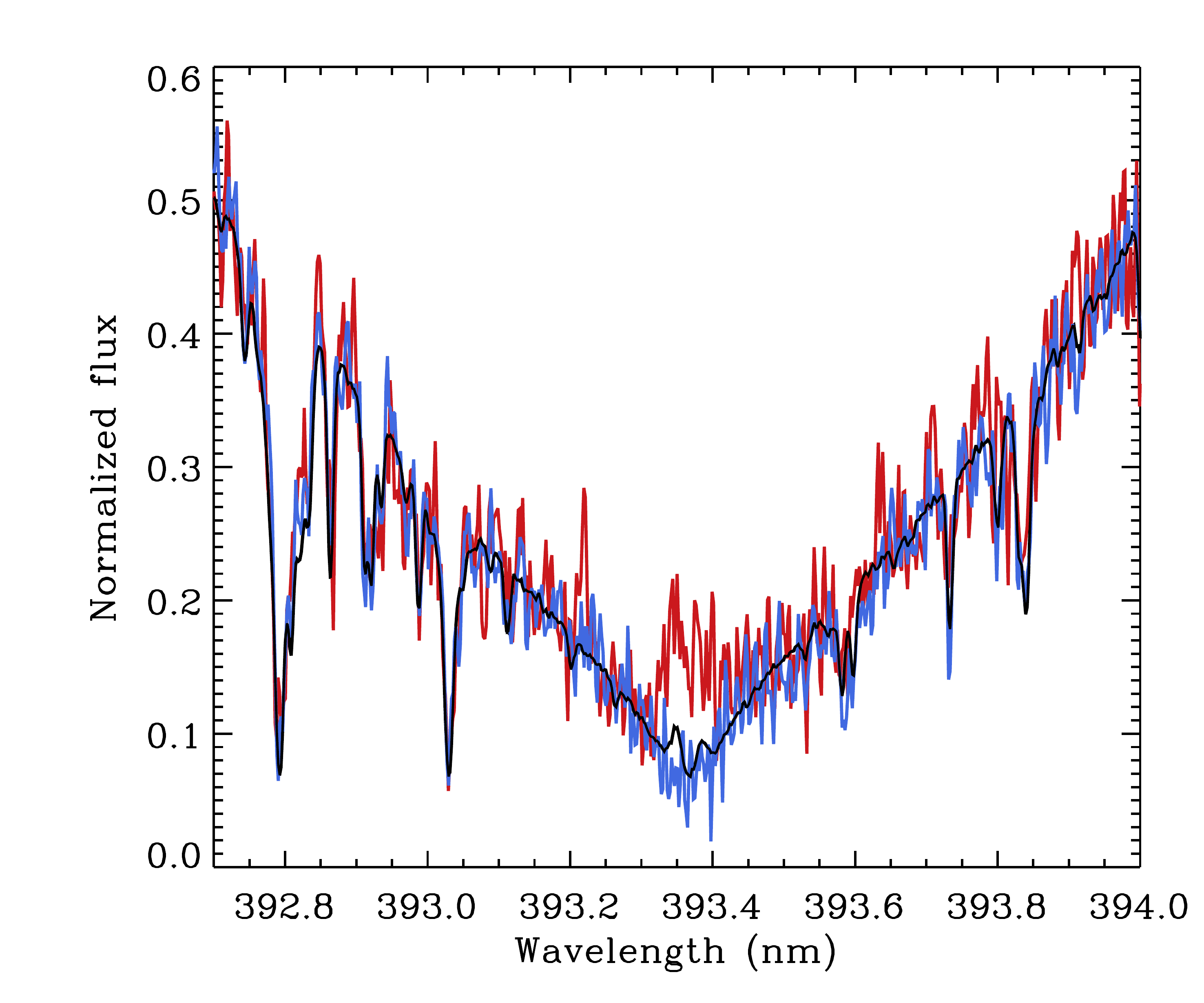} 
\caption{
Comparison of the solar spectrum (black) to two seismic solar analogs, KIC\,4914923 (blue) and KIC\,5774694 (red), around the Ca\,K line observed with the \Hermes spectrograph and illustrative to stars with different magnetic activity levels and ages.
\label{fig:hermes_spec} }
\end{center}
\end{figure}

The chromospheric activity is typically quantified through the classical \Ssymbol~index, as defined by \citet{wilson78}. This formalism measures the strength of the plasma emission in the cores of the \ion{Ca}{II}\,H and K lines in the near ultra violet (UV). The result is dependent on the instrumental resolution but also on the spectral type of the star. We note however that the selected sample of stars was chosen for all having comparable stellar properties to the Sun. The estimated values of the \Ssymbol~index can be thus safely compared between each others and with the Sun as well.
As an illustration, Fig.~\ref{fig:hermes_spec} shows the differences in the Ca\,K line in the case of two stars with different levels of magnetic activity and ages. The represented solar analogs observed with the \Hermes spectrograph are the young and active KIC\,5774694 (red), and the more evolved, less active KIC\,4914923 (blue). The solar spectrum obtained with the \hermes instrument in April 2015 \citep{beck15} is also shown for comparison (black).
Furthermore,  from observations collected with the \Hermes spectrograph of G2-type stars present in the catalog of the Mount Wilson Observatory \citep[MWO,][]{duncan91}, \cite{beck15} determined the instrumental factor to scale the \Ssymbol~index derived with \Hermes into the MWO system to be $\alpha = 23\pm2$.

% -----------------------------------------------------------------------------------------
% -----------------------------------------------------------------------------------------

The values of the \Ssymbol~index measured for the 18 seismic solar analogs and converted into the MWO system are reported in Table~\ref{table:sph}. For each star, the \Ssymbol~index was obtained by taking the mean value of the \Ssymbol~indices calculated for each unnormalized individual spectrum, while the quoted uncertainties correspond to their associated dispersion. Except for KIC\,3241581 ($\Delta$T\,=\,504\,days) and KIC\,10644253 ($\Delta$T\,=\,180\,days), the \Hermes observations analyzed here cover a period $\Delta$T of about 35 days over two observing runs (see Section~\ref{sec:hermes_obs}). Over such short interval of time, the scatter between spectra will be mainly related to the instrumental systematics and the weather conditions, while the possible signature of temporal magnetic variations would remain smaller. 
We note also that in the case of two stars, only one spectrum could be used and that in consequence no dispersion on their \Ssymbol~index is quoted in Table~\ref{table:sph}.
Furthermore, to characterize the quality of the individual observations, the signal-to-noise ratio in the near UV, S/N(Ca), was estimated for each spectrum at the centre of the 90$^\mathrm{th}$ \'echelle order at about 400\,nm. The mean values of S/N(Ca) are also given in Table\,\ref{table:sph}.
We found that the 5 stars with an observed S/N(Ca)\,<\,15 show larger dispersions of their measured \Ssymbol~index, unlike the 13 stars which higher S/N(Ca).
This empirical threshold in S/N(Ca) provides the limit for which the observational and stellar noise dominate the outcome of the \Ssymbol formalism.

% -----------------------------------------------------------------------------------------
\begin{table}[t]
\caption{Comparison of the \sindex values with the ones measured by \citet{karoff13} and \citet{isaacson10}. }
\label{table:table_karoff}
\begin{center}
\begin{tabular}{c c c c}
\hline\hline
KIC	& (1) & (2) & This work\\
\hline
4914923	& $0.137\pm0.005$	& $-$ &	$0.168\pm0.014$\\
6116048	& $0.152\pm0.001$	& 0.157 & $0.157\pm0.001$\\
9098294	& $0.150\pm0.003$	& $-$ & $0.203\pm0.031$\\
\hline
\end{tabular}
\tablebib{(1) \citet{karoff13}; (2) \citet{isaacson10}.}
\end{center}
\end{table}
% -----------------------------------------------------------------------------------------

% -----------------------------------------------------------------------------------------
\subsection{Comparison with previous measurements}
Three stars of our {\it Kepler} sample were observed with the Nordic Optical Telescope (NOT) over the period 2010--2012 and their \Ssymbol~index measured by \citet{karoff13}.  One of these stars was also measured in 2010 as part of the sample by the California Planet Search program \citep{isaacson10}. The corresponding values are given in Table~\ref{table:table_karoff} and compared to the ones measured in this work with the \Hermes spectrograph in June--July 2015. However, we need to keep in mind that the comparison of the \Ssymbol~index derived from different spectrographs is affected by the associated observational, instrumental, and methodological uncertainties as well as by the corresponding derived instrumental scale factor. 
For KIC\,4914923 and KIC\,9098294, the two candidates for a binary system detected in this work (see Table~\ref{table:rv}), the values of the \Ssymbol~index reported by \citet{karoff13} are smaller than the ones measured here, which would indicate an increase of their magnetic activity in the last few years of about 23\% and 35\% respectively. The case of the system KIC\,9098294 is particularly intriguing because a relatively large variation in radial velocity ($\Delta$RV\,=\,41.35\,km\,s$^{-1}$) was measured here within a short period of time (35~days), which is comparable to the variation found in the set of eccentric binary systems studied with the \Hermes spectrograph by \citet{beck14}. At the current moment and without the knowledge of the orbital period and the eccentricity of this system, it can only be speculated if the observed variation in the \Ssymbol~index between the measurements from \citet{karoff13} and this study is connected and modulated with the binarity. Further monitoring is thus needed.
On the other hand, the activity of  KIC\,6116048 seems to have remained stable, as consistent \sindex values are obtained for the three measurements taken five years apart.

%------------------------------------------------------------------
\section{Discussion}
\label{sec:discussion}
\subsection{Comparison between $S_\mathrm{ph}$ and \Ssymbol~index}
The photospheric $S_\mathrm{ph}$ activity proxy can be easily and quickly calculated from existing space photometric observations for a large number of stars observed simultaneously. Moreover, the available observations offer a length of time sufficient enough to cover periods longer than five stellar rotation necessary to measure the $S_\mathrm{ph}$. The estimation of a reliable value of the chromospheric \Ssymbol~index is however more complex as it requires an important investment in telescope time to acquire enough ground-based spectroscopic observations for each individual target.
Furthermore, this is more complicated for these rather faint photometric targets. Additionally, these two proxies measure activity in different stellar regions. It is thus important to understand the relation between the \Ssymbol~index and the $S_\mathrm{ph}$ by comparing the behavior of these two proxies for a common set of stars. 

\begin{figure}[tbp]
\begin{center} 
\includegraphics[width=0.49\textwidth,angle=0]{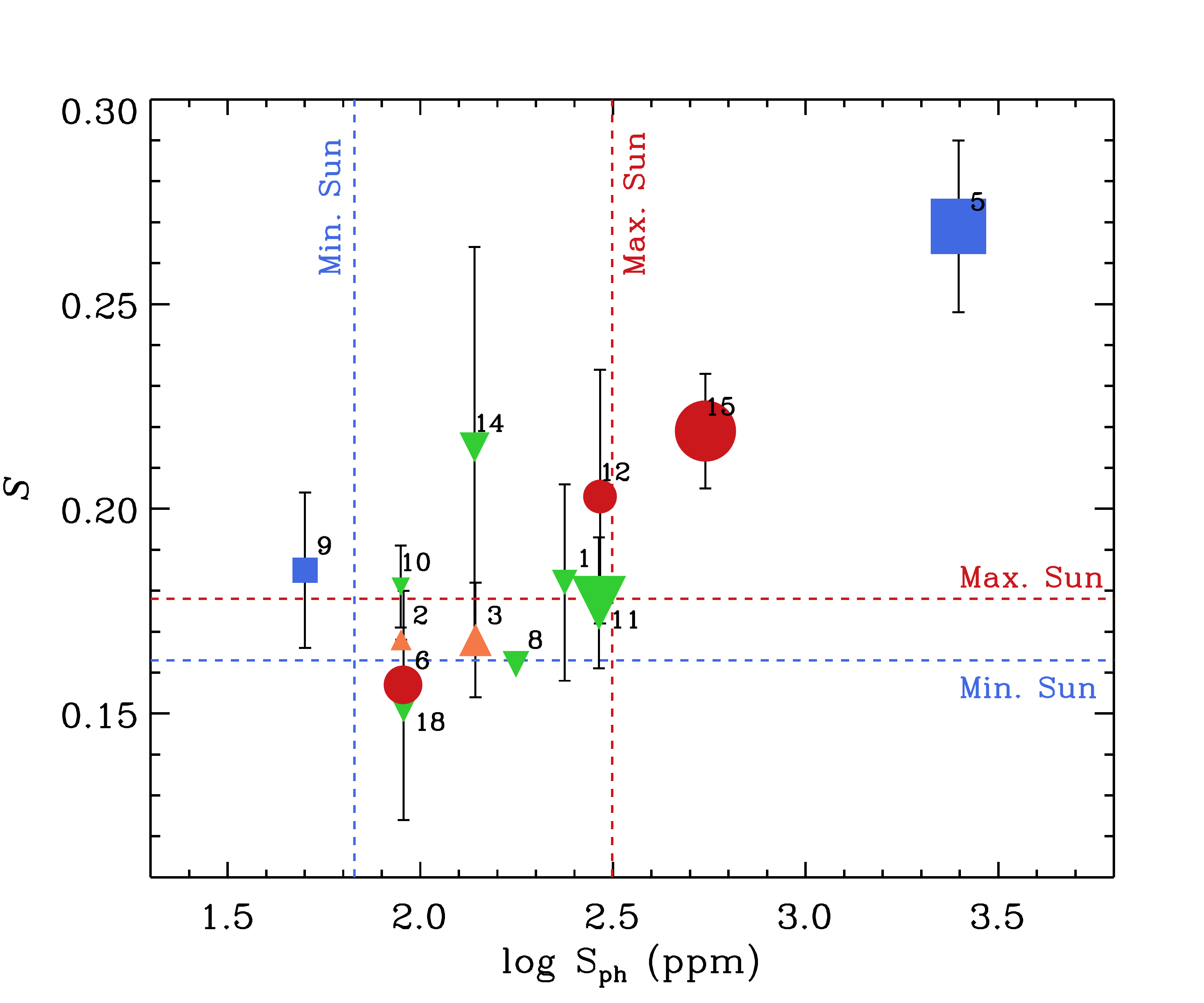}
\end{center}
\caption{\label{fig:hermes}  
Chromospheric \Ssymbol~index derived from the \Hermes observations and calibrated into the MWO system as a function of the photospheric $S_{\text{ph}}$ (in ppm) of the seismic solar analogs observed with the {\it Kepler} satellite. The corresponding activity levels of the Sun at minimum and maximum of its 11-year magnetic cycle are represented by the horizontal and vertical dashed lines. The size of the symbols is inversely proportional to the rotation period. Each star is referred by the same number as given in Tables~\ref{table:prop} and \ref{table:sph}.}
\end{figure}

The comparison between the photospheric $S_\mathrm{ph}$ and chromospheric \Ssymbol-index magnetic activity proxies derived from the {\it Kepler} and \Hermes observations respectively is shown on Fig.~\ref{fig:hermes} for the subset of 13 solar analogs with a S/N(Ca)\,>\,15 in the spectroscopic data (see Section~\ref{sec:sindex}). 
The same color and symbol codes as in Fig.~\ref{fig:kepler} were used, while the size of the symbols is inversely proportional to the rotation period from \citet{garcia14}. Faster rotators are then represented with bigger symbols.
The mean values at minimum and maximum along the solar cycle of the photospheric (Salabert et al., in prep.) and chromospheric \citep[Egeland et al., in prep.;][]{hall04} magnetic activity levels are represented respectively by the vertical and horizontal dashed lines. The resulting activity box corresponds to the range of change in solar activity along the 11-year magnetic cycle. 
Although the sample of stars is small, Fig.~\ref{fig:hermes} indicates that, within the errors, the $S_\mathrm{ph}$ and \Ssymbol indices are complementary. We note also that the $S_\mathrm{ph}$ and \Ssymbol~index proxies were not estimated from contemporaneous {\it Kepler} and \Hermes observations, introducing a dispersion related to possible temporal variations in the stellar activity. Furthermore, the $S_\mathrm{ph}$ proxy corresponds to a mean activity level averaged over a long period of time -- 1422\,days of the {\it Kepler} observations between 2009 and 2013 --, while the \Sindex proxy measures the stellar activity over a shorter temporal snapshot. In addition, we recall that the $S_\mathrm{ph}$ is dependent on the inclination angle and represents a lower limit of the photospheric activity. However, it confirms that the $S_\mathrm{ph}$ can accompany the classical \Ssymbol~index for activity studies.

% -----------------------------------------------------------------------------------------
% -----------------------------------------------------------------------------------------

% -----------------------------------------------------------------------------------------
\begin{figure*}[tbp]
\begin{center} 
\includegraphics[width=0.49\textwidth,angle=0]{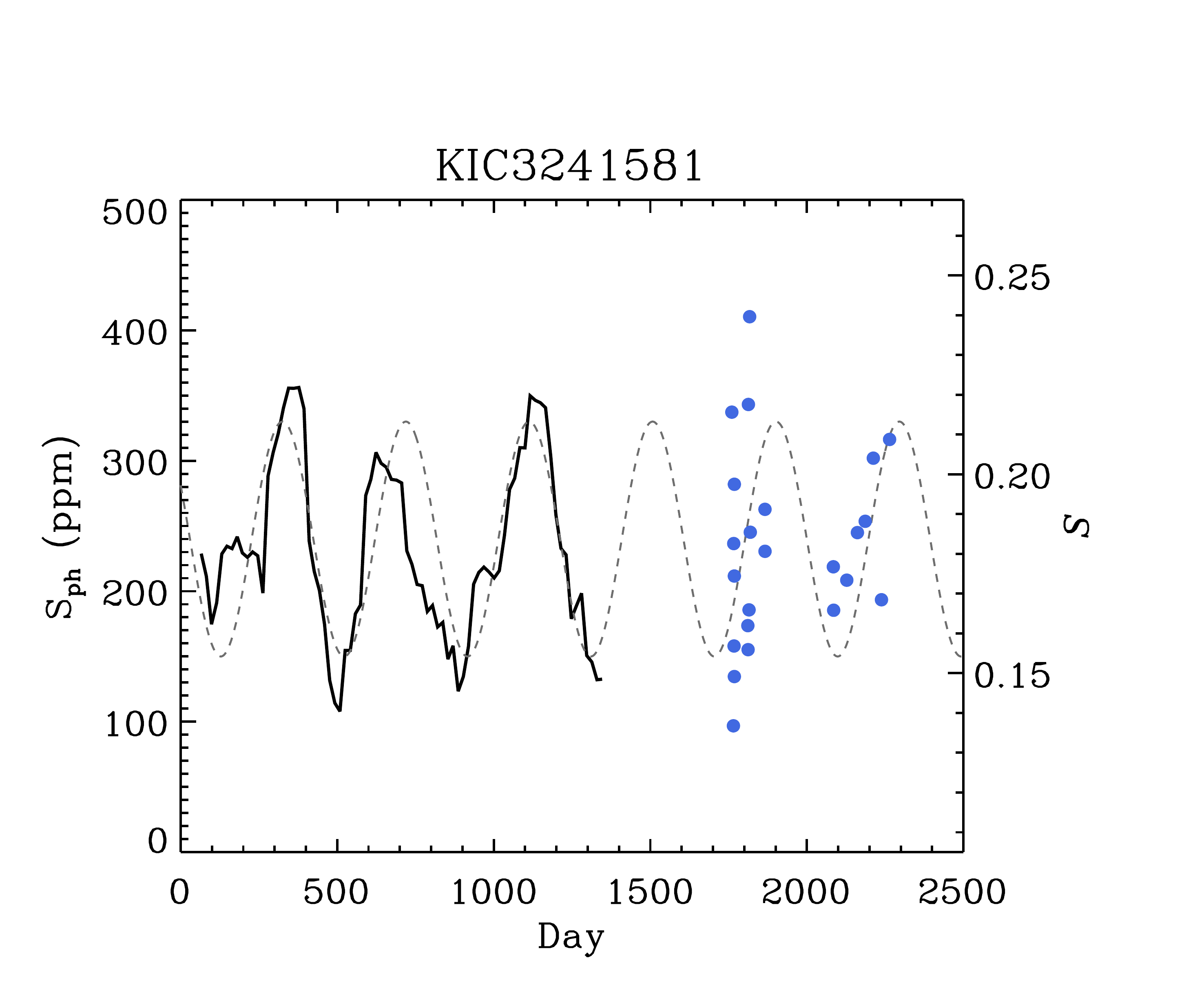}
\includegraphics[width=0.49\textwidth,angle=0]{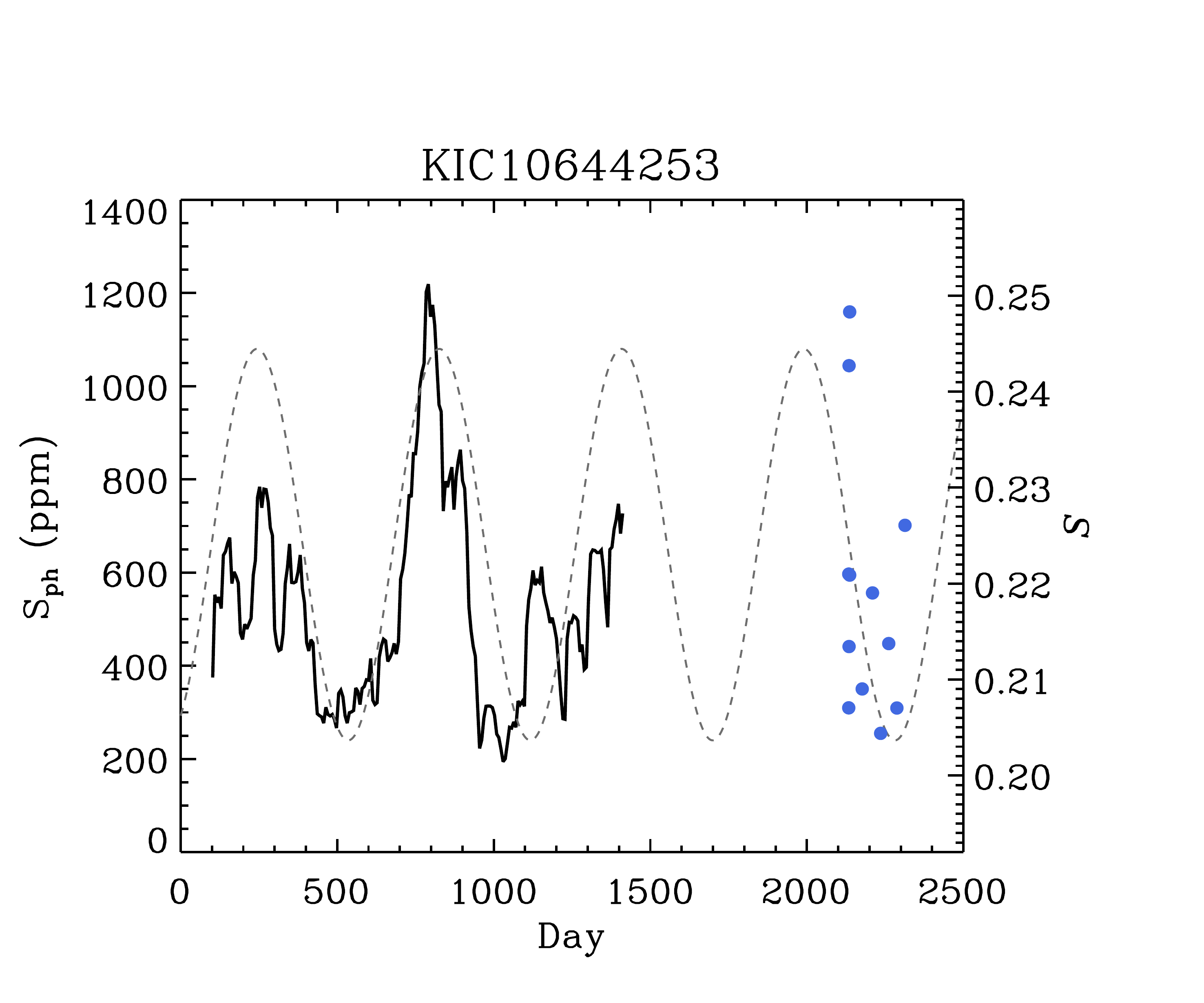}
\end{center}
\caption{\label{fig:twiny} 
(Left panel) Photospheric magnetic activity proxy $S_\mathrm{ph}$ (in ppm, solid black line) of KIC\,3241581 estimated from 1422~days of {\it Kepler} observations as a function of time.  The individual measurements of the \Ssymbol~index obtained with the \Hermes observations few years after the end of the {\it Kepler} mission are represented by the dots.  The gray dashed line represents a sinusoid calculated using the main periodicity found in the $S_\mathrm{ph}$ by a Lomb-Scargle analysis and centered around the first modulation of the  $S_\mathrm{ph}$.
(Right panel) Same as the left panel but for KIC\,10644253. }
\end{figure*} 
% -----------------------------------------------------------------------------------------

% -----------------------------------------------------------------------------------------
\subsection{Temporal variability in KIC\,3241581 and KIC\,10644253}
\label{sec:sindex_variab}
In the case of KIC\,3241581 and KIC\,10644253, for which the time interval, $\Delta$T, of the \Hermes observations cover several months, we checked for any possible temporal variations of the \Ssymbol~index which could be related with variations measured in the $S_\mathrm{ph}$.
For these two stars, the photospheric $S_\mathrm{ph}$ shows a clear modulation as a function of time (Fig.~\ref{fig:twiny}) and the associated Lomb-Scargle periodograms return a main period of about 394~days for  KIC\,3241581 and of 582~days for KIC\,10644253, the latter being already discussed in \citet{salabert16} along comparable variations in the p-mode frequency shifts. 
The $\sim$\,1.1-year variations in KIC\,3241581 at a rotation period comparable to the Sun may be analogous to the solar quasi-biennial oscillation observed in various activity proxies \citep[see, e.g.,][and references therein]{bazi14}. Although the associated modulation of about 200\,ppm is relatively large, it is however close to the {\it Kepler} orbital period. Even though that seems unlikely, we cannot rule out yet any pollution  related to the {\it Kepler} orbit.  
The $\sim$\,1.6-year variations in the young 1-Gyr-old KIC\,10644253 at a rotation period $\sim$\,11~days previously observed by \citet{salabert16} is analogous to what is found by \citet{egeland15} in the Mount Wilson star HD\,30495, having very close stellar properties and falling on the inactive branch reported by \citet{bohm07}. 

The associated individual measurements of the \Ssymbol~index obtained with the \Hermes observations few years after the end of the {\it Kepler} mission are represented by the dots on Fig.~\ref{fig:twiny}. For both stars, we have also overplotted a sine function calculated using the periodicities given by the Lomb-Scargle analysis above and centered around the first modulation of the  $S_\mathrm{ph}$. Although this is only qualitative 
and that it could be related to the intrinsic \Hermes instrumental and observational scatter, the \sindex values for KIC\,3241581 show however striking variations which appear to be in close temporal phase with the $S_\mathrm{ph}$ during two consecutive periods of increasing activity. Only additional spectroscopic observations could confirm these variations in the \Ssymbol~index, mainly during periods associated to a possible decreasing phase of activity based on $S_\mathrm{ph}$. For KIC\,10644253, a longer time interval is required because the main periodicity measured in the $S_\mathrm{ph}$ is larger than the \Hermes observations analyzed here.

% -----------------------------------------------------------------------------------------
\begin{figure}[tbp]
\begin{center} 
\includegraphics[width=0.49\textwidth,angle=0]{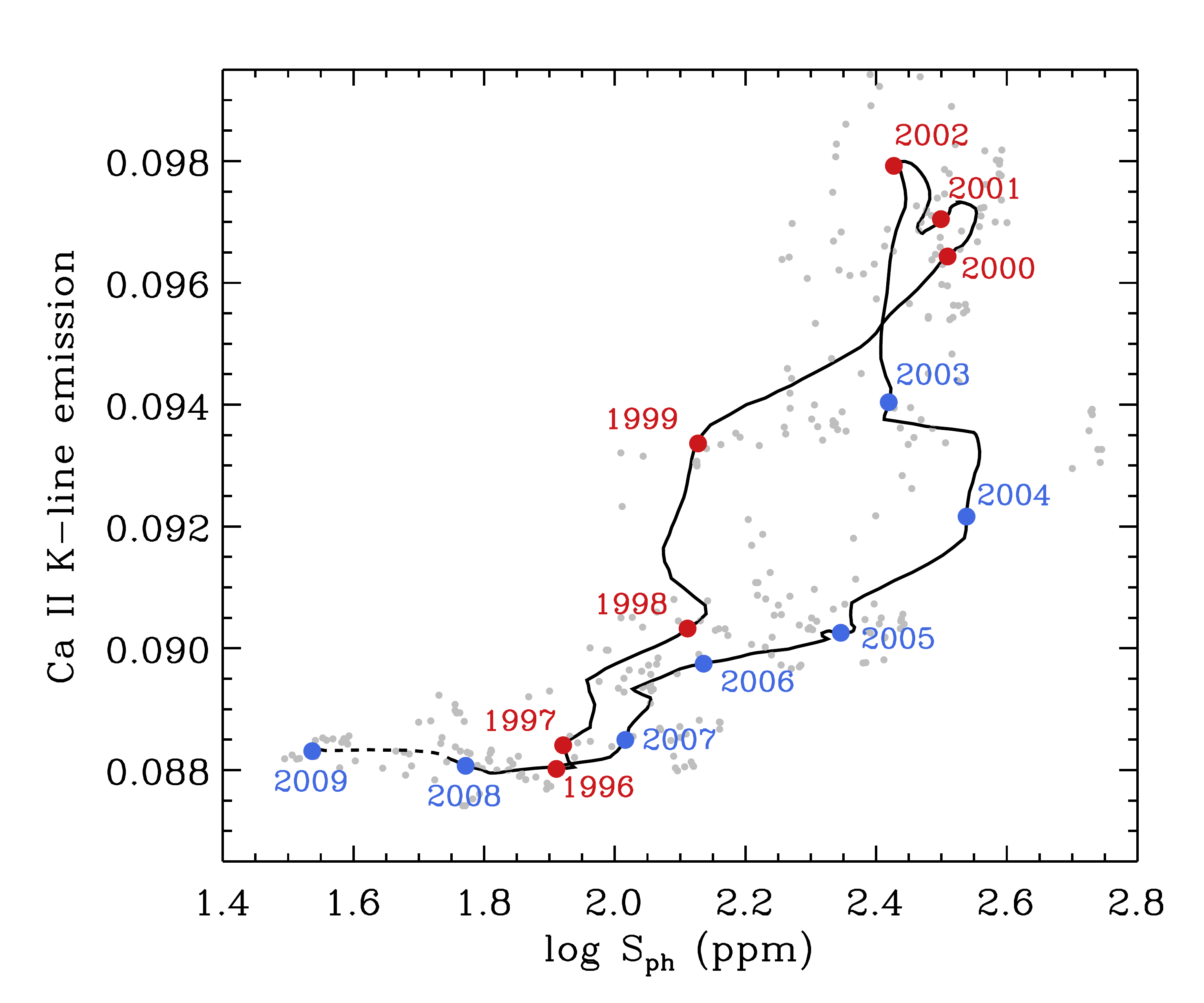}
\end{center}
\caption{\label{fig:hyste} 
Solar Ca\,K-line emission as a function of the photospheric magnetic proxy, $S_\mathrm{ph,\,\sun}$ (in ppm), of the Sun during the solar cycle 23 between 1996 and 2008 (solid line) and smoothed over 1 year. The first year of cycle 24 is also represented in dashed line. The start of each year is indicated: red for the rising phase and blue for the falling phase of the activity cycle. The individual data points before smoothing are shown in gray.}
\end{figure} 
% -----------------------------------------------------------------------------------------

% -----------------------------------------------------------------------------------------
\subsection{The case of the Sun}
To better understand the relation between the photospheric and chromospheric magnetic activity of the solar analogs represented in Fig.~\ref{fig:hermes}, we compared the temporal variations of both quantities in the case of the Sun along its 11-year activity cycle.
Unfortunately, no publicly long-term monitoring of the solar \Ssymbol~index is available. Several scaling relationships were derived to convert the chromospheric Ca K-line emission index of the Sun into \Ssymbol~index \citep[e.g.,][]{duncan91,white92,radick98}, but they provide different estimates of the solar minima and cycle amplitudes (within 10\% and 4\% respectively). We thus decided to compare the photospheric $S_\mathrm{ph,\,\sun}$ proxy of the Sun directly to the Ca K-line emission index measured at Sacramento Peak/National Solar Observatory facility\footnote{The data from the Ca K-line monitoring program are available at \url{nsosp.nso.edu/node/15}.}. Figure~\ref{fig:hyste} shows the solar Ca K-line emission index as a function on a x-log scale of the corresponding $S_\mathrm{ph,\,\sun}$ calculated from the photometric VIRGO/SPM observations. The period covered corresponds to the entire solar cycle 23 (1996\,--\,2008) as in Sec.~\ref{sec:sph} and Figs.~\ref{fig:kepler} and \ref{fig:hermes}. The nearly daily measurements of Ca K were averaged over the same subseries used to estimate the solar $S_\mathrm{ph,\,\sun}$ (i.e. $5\,\times\,P_\mathrm{rot,\,\sun}\,=\,125$\,days with $P_\mathrm{rot,\,\sun}\,=\,25$\,days). 
For illustrative purpose on Fig.~\ref{fig:hyste}, the data points were smoothed over a period of 1\,year. The individual data points before smoothing are shown in light gray. The start of each year is also indicated. 
The relationship between the two proxies varies between the rising and falling phases of the solar cycle following an hysteresis pattern.
Moreover, a saturation in chromospheric activity is observable during the unusual long and deep activity minimum of cycle~23, which could be associated to the basal chromospheric component as observed in inactive stars and in the quiet Sun \citep[e.g.,][]{schrijver89,schroder12,stenflo12}. However, this saturation is interestingly not visible in the photospheric activity of the Sun, $S_\mathrm{ph,\,\sun}$. Nonetheless, such hysteresis has been observed in a wide range of solar observations between photospheric and chromospheric activity proxies, as well as with the p-mode frequency shifts \citep[e.g.][]{bach94,chano98,tripathy01,ozguc12}. Finally, Fig.~\ref{fig:hyste} supports the complementarity between the photospheric, $S_\mathrm{ph}$, and chromospheric, \Ssymbol~index, proxies observed for the solar analogs in Fig.~\ref{fig:hermes}.

% -----------------------------------------------------------------------------------------
\subsection{Comparison with Mount Wilson FGK stars}
\begin{figure}[tbp]
\begin{center} 
\includegraphics[width=0.49\textwidth,angle=0]{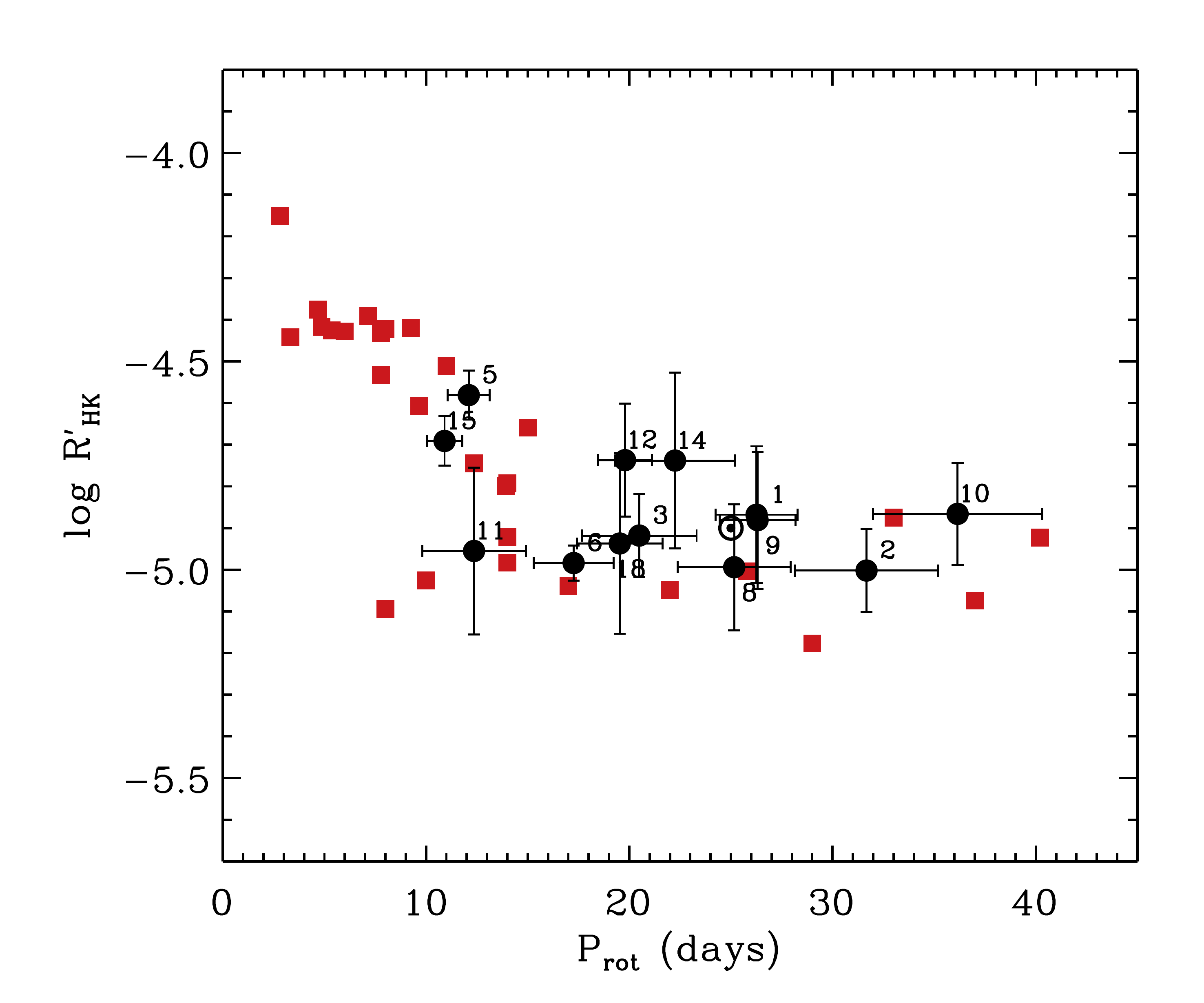}
\end{center}
\caption{\label{fig:rhk} 
Chromospheric flux ratio, log\,$R'_\mathrm{HK}$, as a function of the rotation period, $P_\mathrm{rot}$ (in days). The black dots correspond to the {\it Kepler} seismic solar analogs, and the red squares to the solar-analog stars found in the MWO sample of FGK stars. Each represented {\it Kepler} star is referred by the same number as given in Tables~\ref{table:prop} and \ref{table:sph}. The position of the Sun is also indicated with its astronomical symbol for a log\,$R'_\mathrm{HK}=-4.96$ \citep{hall09} and a rotation of 25~days.}
\end{figure}

Furthermore, in order to place our sample into context, we assembled a catalog of solar-analog stars with measured \ion{Ca}{II}\,H and K activity and rotation from the FGK stars observed with the Mount Wilson program \citep{baliunas96,donahue96}. In case where measurements were available in both works, values from \citet{donahue96} were used. These works give activity averages from synoptic time series lasting up to 30~years. The values of \Ssymbol from our \Kepler sample and the MWO stars were converted into the chromospheric emission fraction, $R'_{\rm HK}$, using the prescription of \citet{noyes84}, based on Tycho-2 $(B-V)$ measurements (see Table~\ref{table:sph}). The associated uncertainties were derived using standard error propagation methods. The index, $R'_{\rm HK}$, removes the photospheric contribution to the \ion{Ca}{II} H and K bands. The Tycho-2 photometry catalogue was obtained from Hipparcos by \citet{hog00}. A rough selection of solar-analog stars was made by transforming the published $(B-V)$ color index into effective temperature using the scaling relation from \citet[][Eq. 2]{noyes84}: $\log T_{\rm eff} = 3.908 - 0.234\times(B-V)$, and applying the same selection criterium in effective temperature used for the \emph{Kepler} sample, i.e., 5520\,K $\leq T_\mathrm{eff} \leq$ 6030\,K. The final catalog contains 28 MWO solar analogs. 
Figure \ref{fig:rhk} shows the $R'_{\rm HK}$ activity proxy of the MWO and \Kepler solar analogs as a function of their rotation periods. 
The rotational modulation of the MWO stars was measured directly from their \sindex time series using a Lomb-Scargle periodogram on bins containing seasonal observations from 150--200 days in length \citep{donahue96}. Only the subset of the 13 {\it Kepler} solar analogs with a spectroscopic S/N(Ca)\,>\,15 (see Section~\ref{sec:sindex}) is represented.

Our {\it Kepler} sample is observed to be in the same vicinity as the MWO sample, which gives additional confidence in our determination of the activity. The MWO stars also show a rough trend of higher activity for faster rotation: unfortunately our \emph{Kepler} sample lacks of stars rotating faster than 10~days.

% -----------------------------------------------------------------------------------------
% -----------------------------------------------------------------------------------------

% -------------------------------------------------------------------------
\section{Conclusions}
\label{sec:conclusion}
The study of the characteristics of the surface activity of solar analogs can provide new constraints in order to better understand the magnetic variability of the Sun, and its underlying dynamo processes, along of its evolution compared to other stars.
We thus analyzed here the sample of main-sequence stars observed by the {\it Kepler} satellite for which solar-like oscillations were detected \citep{chaplin14} and rotational periods measured \citep{garcia14}. Using published stellar parameters derived from the global or the individual asteroseismic properties \citep{mathur12,chaplin14,metcalfe14}, we identified 18 seismic solar analogs rotating between 10 and 40~days.
We then studied the properties of the photospheric and chromospheric magnetic activity of these stars in relation of the Sun. The photospheric activity proxy, $S_\mathrm{ph}$, was derived by means of the analysis of the {\it Kepler} observations collected over 1422~days of science operations (from 2009 June 20 to 2013 May 11). The chromospheric activity proxy, \Ssymbol~index, was measured with follow-up, ground-based  spectroscopic observations in 2014 and 2015 with the \Hermes instrument mounted to the 1.2\,m \textsc{Mercator} telescope in La Palma (Canary Islands, Spain).

Although the size of the sample is rather small, it constitutes the most suitable set of seismic solar analogs which can be identified today with {\it Kepler}. In the future, the Transiting Exoplanet Survey Satellite \citep[TESS;][]{ricker15} and PLATO \citep{rauer14} space missions will likely expand this list. Nevertheless, we showed that the magnetic activity of the Sun is comparable to the activity of the seismic solar analogs studied here, within the maximum-to-minimum activity variations of the Sun during the 11-year cycle.
As expected, the youngest and fastest rotating stars are observed to actually be the most active of our {\it Kepler} sample, while the activity of stars older than the Sun seems to not evolve much with age.
 
Furthermore, the comparison of the photospheric $S_\mathrm{ph}$ with the well-established chromospheric \Ssymbol~index shows that the $S_\mathrm{ph}$ index can be used to provide a suitable magnetic activity proxy. Moreover, it can be easily estimated for a large number of stars with known surface rotation observed simultaneously with photometric space missions whose durations cover periods of time longer than five stellar rotation needed to measure the Sph. On the other hand, the estimation of the associated \Ssymbol~index would be highly difficult to achieve as it would require a lot of time of ground-based telescopes to collect enough spectroscopic data for each individual target. Moreover, these photometric targets are rather faint for spectroscopy observations. Nevertheless, we need to keep in mind that such a photospheric proxy provides a lower limit of the stellar activity because it is dependent on the inclination angle of the rotation axis in respect to the line of sight. This is assuming as well that the starspots are formed over comparable ranges of latitude as in the Sun. Accurate estimates of the inclination angles will be then of great importance in order to provide tighter constraints on the stellar magnetic activity for gyro-magnetochronology studies.

% -----------------------------------------------------------------------------------------

% -----------------------------------------------------------------------------------------
\begin{acknowledgements}
 The authors wish to thank the entire {\it Kepler} team, without whom these results would not be possible. Funding for this Discovery mission is provided by NASA Science Mission Directorate. The ground-based observations are based on spectroscopy made with the Mercator Telescope, operated on the island of La Palma by the Flemish Community, at the Spanish Observatorio del Roque de los Muchachos of the Instituto de Astrof\'isica de Canarias. This work utilizes data from the National Solar Observatory/Sacramento Peak \ion{Ca}{ii} K-line Monitoring Program, managed by the National Solar Observatory, which is operated by the Association of Universities for Research in Astronomy (AURA), Inc. under a cooperative agreement with the National Science Foundation.
The research leading to these results has received funding from the European Community's Seventh Framework Program ([FP7/2007-2013]) under grant agreement no. 312844 (SPACEINN) and under grant agreement  no.  269194  (IRSES/ASK). DS and RAG acknowledge the financial support from the CNES GOLF and PLATO grants.  
PGB acknowledges the ANR (Agence Nationale de la Recherche, France) program IDEE (n$\degr$ ANR-12-BS05-0008) "Interaction Des Etoiles et des Exoplan\`etes". 
RE is supported by the Newkirk Fellowship at the High Altitude Observatory.
SM acknowledges support from the NASA grants NNX12AE17G and NNX15AF13G. TM acknowledges support from the NASA grant NNX15AF13G.
JDNJr acknowledges support from CNPq PQ 308830/2012-1 CNPq PDE Harvard grant.
DS acknowledges the Observatoire de la C\^ote d'Azur for support during his stays. This  research  has  made use of the
SIMBAD database, operated at CDS, Strasbourg, France.
\end{acknowledgements}	
%-------------------------------------------------------------------

% -----------------------------------------------------------------------------------------

% -----------------------------------------------------------------------------------------

% -----------------------------------------------------------------------------------------
\begin{appendix}
\section{Filter impact on the estimated $S_\mathrm{ph}$}
\label{app:1}
\begin{figure*}[tbp]
\begin{center} 
\includegraphics[width=0.45\textwidth,angle=0]{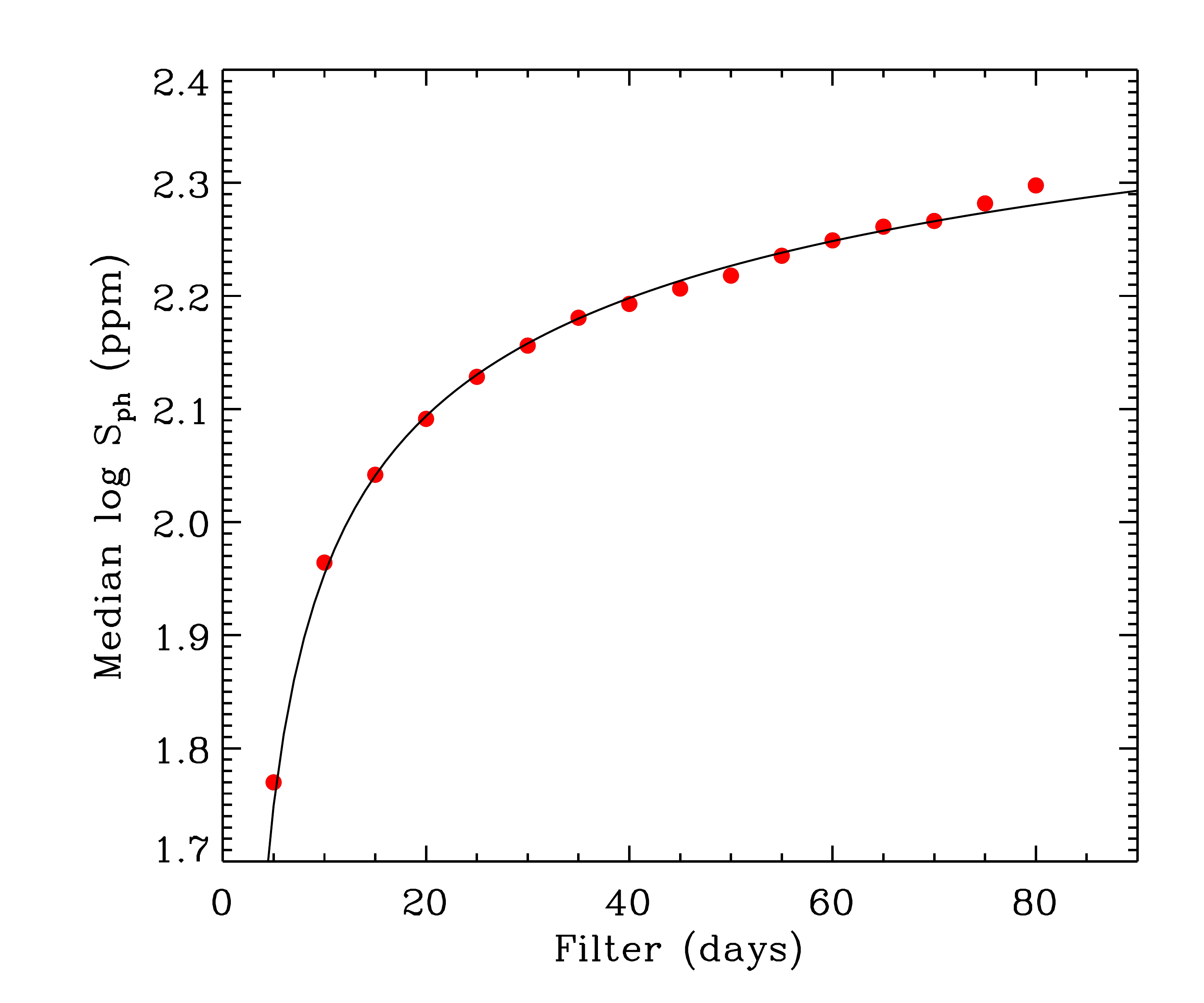}
\includegraphics[width=0.45\textwidth,angle=0]{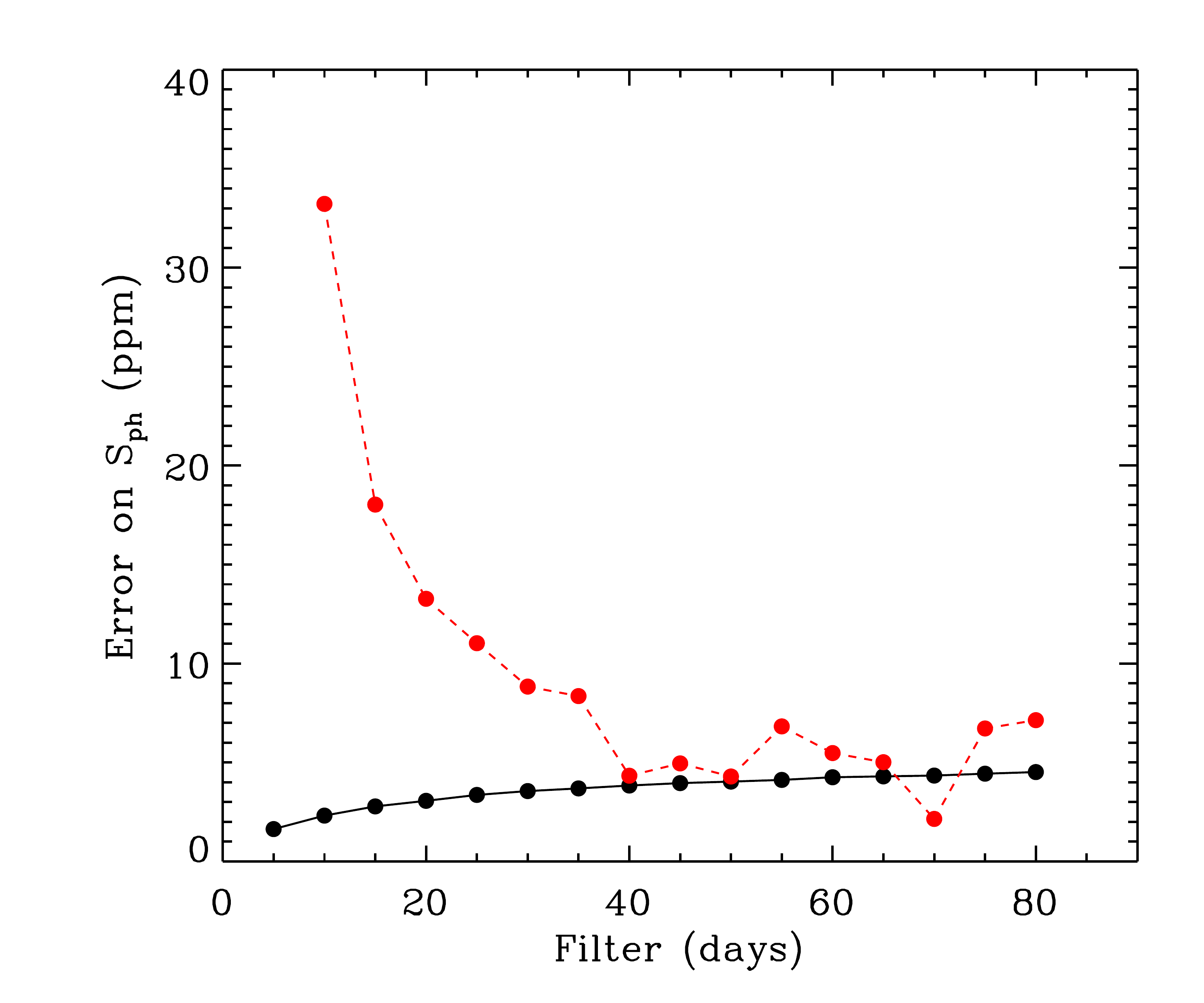}
\end{center}
\caption{\label{fig:filt}  
(Left panel) Median values of the photospheric activity proxy, $S_\mathrm{ph}$ (in ppm), calculated for the 310 {\it Kepler} solar-like stars for which rotation periods were measured \citep{garcia14} as a function of the length in days of the applied filter $\mathbb{F}$ (red dots).  The solid line corresponds to the associated fit from Eq.~\ref{eq:eq1}. (Right panel) Median values of the associated errors (in ppm) of the estimated $S_\mathrm{ph}$ of the 310 {\it Kepler} solar-like stars as a function of the filter $\mathbb{N}_\mathrm{day}$ (solid line). The red dashed-line corresponds to the difference in $S_\mathrm{ph}$ between consecutive values of the filter $\mathbb{N}_\mathrm{day}$.
}
\end{figure*}

The data processing of the photometric {\it Kepler} observations can have important impacts on the estimated values of $S_\mathrm{ph}$.
Indeed, the choice of the high-pass filter applied to the original {\it Kepler} data throughout the calibration procedure affects the returned estimates of $S_\mathrm{ph}$. For a given star, the $S_\mathrm{ph}$ will get smaller because the filter bandpass gets narrower as represented on the left panel of Fig.\ref{fig:filt}. It shows the median values of $S_\mathrm{ph}$ calculated for the 310 solar-like stars for which rotation periods, $P_\mathrm{rot}$, were measured by \citet{garcia14} for different lengths of filter, $\mathbb{N}_\mathrm{day}$, applied during the data processing, going from 5 days to 80 days, with 5-day increments. Such dependence for the solar-like oscillating stars was evaluated to be well described by the following formulation:

\begin{equation}
\log (S_\mathrm{ph}) = 4.25-\frac{2.29}{\log (\mathbb{N}_\mathrm{day})^{0.24}},
\label{eq:eq1}
\end{equation}
where $\mathbb{N}_\mathrm{day}$ represents the number of days of the applied filter.
The right panel of Fig.~\ref{fig:filt} shows the median values of the associated errors of the estimated $S_\mathrm{ph}$ of the 310 solar-like stars as a function of the filter. For comparison, the red dashed-line shows the difference in $S_\mathrm{ph}$ between consecutive values of the filter as an indicator of the error introduced by a change in the filter bandpass. For filters 40-day wide upwards, the impact of a change in filter appears to be within the errors on $S_\mathrm{ph}$.
We decided then to use a filter of $\mathbb{N}_\mathrm{day}$\,=\,55~days as the rotation periods of our sample of stars are faster than 40~days. 

% -----------------------------------------------------------------------------------------
% -----------------------------------------------------------------------------------------
% -----------------------------------------------------------------------------------------

\end{appendix}
\end{document}